\documentclass[a4paper,fleqn,useAMS,usenatbib]{mnras}

\usepackage[T1]{fontenc}
\usepackage{ae,aecompl} 

\usepackage{graphicx}
\usepackage{amssymb}
\usepackage{amsfonts}
\usepackage[usenames,dvipsnames]{color}

\usepackage{lscape}

\def\integral{{\it INTEGRAL}}
\def\integrale{{\it INTEGRAL }}
\def\xte{{\it RXTE}}
\def\xtee{{\it RXTE }}
\def\gro{{\it CGRO}}
\def\groe{{\it CGRO }}

\def\gingae{{\it GINGA }}

\def\swift{{\it Swift}}
\def\swifte{{\it Swift }}
\def\suzaku{{\it Suzaku}}
\def\suzakue{{\it Suzaku }}

\def\ascae{{\it ASCA }}
\def\sax{{\it BeppoSAX}}
\def\saxe{{\it BeppoSAX }}
\def\xmm{{\it XMM-Newton}}
\def\xmme{{\it XMM-Newton }}
\def\chandra{{\it Chandra}}

\def\nust{{\it NuSTAR}}
\def\nuste{{\it NuSTAR }}

\def\mnras{\rm MNRAS}
\def\apj{\rm ApJ}
\def\apjl{\rm ApJL}
\def\apjs{\rm ApJS}
\def\aj{\rm AJ}
\def\aap{\rm A\&A}
\def\aapr{\rm A\&ARv}

\def\pasj{\rm PASJ}

\def\gca{\rm Geochim. Cosmochim. Acta}
\def\memsai{\rm Mem. S.A.It.}
\def\ssr{\rm SSRv}
\def\nat{\rm Nature}
\title[Hard X-ray spectra of Seyfert galaxies]
{A comprehensive analysis of the hard X-ray spectra of bright Seyfert 
galaxies\thanks{Based on observations with \integral, an ESA project with 
instruments and science data centre funded by ESA member states (especially the 
PI countries: Denmark, France, Germany, Italy, Switzerland, and Spain), the 
Czech Republic, and Poland and with the participation of Russia and the US.}}

\author[P. Lubi\'nski et al.]{P. Lubi\'nski$^{1}$\thanks{E-mail: 
P.Lubinski@if.uz.zgora.pl}, V. Beckmann$^{2}$, L. Gibaud$^{3}$\thanks{present 
address: Observatoire Fran\c{c}ais des Tornades et des Orages Violents},
S. Paltani$^{3}$, I. E. Papadakis$^{4,5}$, C. Ricci$^{6}$, 
\newauthor S. Soldi$^{2}$, M. T\"urler$^{3}$, R. Walter$^{3}$, 
A. A. Zdziarski$^{7}$ \\
$^{1}$Institute of Physics, University of Zielona G\'{o}ra, Licealna 9, 
PL-65-417 Zielona G\'{o}ra, Poland\\
$^{2}$Fran\c{c}ois Arago Centre, APC, Universit\'{e} Paris Diderot, CNRS/IN2P3, 
CEA/Irfu, Observatoire de Paris, 13 rue Watt, 75013 Paris, France\\      
$^{3}$Department of Astronomy, University of Geneva, Chemin d'Ecogia 16, 
CH-1290 Versoix, Switzerland\\
$^{4}$Physics Department, University of Crete, PO Box 2208, 710 03 Heraklion, 
Crete, Greece\\
$^{5}$IESL, Foundation for Research and Technology, 711 10, Heraklion, Crete, 
Greece\\
$^{6}$Pontificia Universidad Cat{\'o}lica de Chile, Instituto de Astrof{\'i}sica, 
Casilla 306, Santiago 22, Chile; EMBIGGEN Anillo, Concepci{\'o}n, Chile\\
$^{7}$Centrum Astronomiczne im. M. Kopernika, Bartycka 18, PL-00-716 Warszawa, 
Poland}

\date{Accepted XXX. Received YYY; in original form ZZZ}

\pubyear{2015}
 
\begin{document}

\label{firstpage}

\pagerange{\pageref{firstpage}--\pageref{lastpage}} 

\maketitle

\begin{abstract}

Hard X-ray spectra of 28 bright Seyfert galaxies observed with \integrale were 
analyzed together with the X-ray spectra from \xmm, \suzakue and \xte. These 
broad-band data were fitted with a model assuming a thermal Comptonization as a 
primary continuum component. We tested several model options through 
a fitting of the Comptonized continuum accompanied by a complex absorption and a 
Compton reflection. Both the large data set used and the model space explored 
allowed us to accurately determine a mean temperature $kT_{\rm e}$ of the 
electron plasma, the Compton parameter $y$ and the Compton reflection strength 
$R$ for the majority of objects in the sample. Our main finding is that a vast 
majority of the sample (20 objects) is characterized by $kT_{\rm e}$ $<$ 
100~keV, and only for two objects we found $kT_{\rm e}$ $>$ 200~keV. The median 
$kT_{\rm e}$ for entire sample is 48$_{-14}^{+57}$~keV. The distribution of the 
$y$ parameter is bimodal, with a broad component centered at $\approx$ 0.8 and 
a narrow peak at $\approx$ 1.1. A complex, dual absorber model improved the fit 
for all data sets, compared to a simple absorption model, reducing the fitted 
strength of Compton reflection by a factor of about 2. Modest reflection (median 
$R$ $\approx$ 0.32) together with a high ratio of Comptonized to seed photon 
fluxes point towards a geometry with a compact hard X-ray emitting region well 
separated from the accretion disc. Our results imply that the template Seyferts 
spectra used in AGN population synthesis models should be revised. 
 
\end{abstract}


\section{Introduction}
\label{intro}

The spectra of Seyfert galaxies in the hard X-ray and soft $\gamma$-ray band 
can be well approximated by a power-law model with a high-energy cut-off
accompanied by a Compton reflection component (for a review see e.g., 
\citealt{Beckmann2012}). Evidence for that was collected already by the \groe 
(e.g., \citealt{Zdziarski2000}) and \saxe satellites (e.g., \citealt{Dadina2007}). 
Since the late 1990s there is a consensus that the main mechanism responsible 
for this emission is the thermal Comptonization of seed photons in the plasma 
cloud surrounding a black hole (BH) in the centre of the AGN (e.g., 
\citealt{Svensson1999}). Unfortunately, due to a limited sensitivity of the 
instruments a detailed quantitative analysis using a physical Comptonization 
model instead of a phenomenological one is not common. The most sensitive soft 
$\gamma$-ray detector so far, OSSE onboard \gro, excluded the presence of a 
strong non-thermal emission in Seyferts \citep{Zdziarski1995}. There remains, 
however, a rather large uncertainty about the typical temperature $kT_{\rm e}$ 
of the electrons forming the plasma cloud and the mean Thomson optical depth  
$\tau$ of this region. The studies based on the data from OSSE and other 
contemporary satellites point towards a low mean temperature $kT_{\rm e}\simeq 
70$--80~keV and a large optical depth $\tau\simeq 1.7$ \citep*{Zdziarski2000}. 
On the other hand, \citet{Petrucci2001} found high values of $kT_{\rm e}\simeq 
170$--320~keV and small $\tau\simeq 0.05$--0.20 for 6 Seyfert type 1 galaxies 
observed by \sax. 

The \integrale \citep{Winkler2003} and \swifte \citep{Gehrels2004} satellites, 
despite the detectors not being as sensitive above 50~keV as OSSE, improved the 
situation thanks to their large field of view resulting in a long total exposure 
time and a better spatial resolution than OSSE and \sax/PDS, important for more 
crowded fields. Two other contemporary hard X-ray satellites, \suzakue 
\citep{Mitsuda2007} and \nuste \citep{Harrison2013}, are suited more for the 
shorter observations of single objects, due to their smaller field of view. 
\nuste hosts the most sensitive detectors in the 10--80~keV band and provides 
high-quality spectra but cannot help with constraining a cut-off energy well 
above 100~keV. Therefore, a comprehensive analysis of average hard 
X-ray spectra for a larger sample of AGN can be performed predominantly with 
the \integrale and \swifte data.   

Reliable estimates of the plasma temperature and optical depth are 
crucial elements for drawing a picture of the central engine of AGN. Another
such element is the knowledge of the system geometry, in particular the extent
of the plasma region and its location relative to the accretion disc. As it was
demonstrated for NGC~4151, even in the case of the bright state of this brightest 
Seyfert galaxy the quality of the spectra does not allow to distinguish which 
geometry option of the Comptonization model provides the best spectral fit 
\citep{Lubinski2010}. Thus, until much better spectra are provided by future
satellites for the plasma geometry studies \citep{Petrucci2008}, additional 
information is needed to constrain the structure of the system. 
The most promising is the Compton reflection component, originating in the 
accretion disc and torus illuminated by the central X-ray source.

Compton reflection was found to be a common feature of Seyfert spectra 
already by the \gingae satellite \citep{Nandra1994}. Further analysis of the
AGN and Galactic BH data revealed that the strength of the Compton reflection
$R$ relative to a reflection from an infinite slab is quite strongly correlated 
with the photon index $\Gamma$ of the X-ray spectra \citep{Zdziarski1999}. 
This correlation together with the X-ray flux/$\Gamma$ correlation
support a scenario with a 
feedback between a hot Comptonizing cloud and a cold reflecting medium in the 
vicinity of the central black hole \citep{Zdziarski2003}. One of the 
possibilities explaining the observed phenomena is the existence of a hot inner 
flow surrounded by a colder accretion disc, consistent with an ADAF geometry 
(e.g., \citealt{Abramowicz1995}, \citealt{Yuan2001}). Due to the presence of a 
second reflector, namely a distant torus or clumpy absorber, the situation for 
AGN is more complex 
than for binary systems and an interpretation of the $\Gamma$-$R$ correlation is 
not straightforward. However, as demonstrated for NGC~4151, good quality data 
for both low- and high-flux states can be used to disentangle the reflection 
from the disc and from the torus \citep{Lubinski2010}.

Information about $kT_{\rm e}$ and $\tau$ of the plasma and $R$ of the 
reflecting medium can be affected by some limitations. First of all, it is 
quite common that the hard X-ray spectra from \sax, \xte, \integral, \suzakue 
and \swifte used for the spectral fit lack a clear detection above 100 keV or 
even below. This makes the analysis of a high-energy cut-off present in both 
primary Comptonized and reflected components less reliable. Thus, for example, 
any result placing $kT_{\rm e}$ well above the upper limit of the spectrum 
appears dubious. Another limitation comes from the spectral models applied. 
Phenomenological models typically assume a power-law with an exponential 
cut-off, which is not as sharp as the cut-off predicted by Comptonization 
models \citep{Zdziarski2003}. This, in turn, can lead to biased results for the 
reflected component, when the primary component has an unphysical form. 

The cosmic X-ray background (CXB) is expected to be the sum of the emission from 
the supermassive BH accreting systems (e.g., \citealt{Mushotzky2000}). Therefore, 
the CXB spectrum, peaked at around 30 keV, provides a strong constraint on the 
cut-off energy of a typical AGN. The AGN population synthesis models reproduce 
the CXB emission assuming that the main constituent of the CXB spectrum below 
200~keV is the accumulated emission from Seyfert galaxies. Template spectra used 
in these models are in the form of the absorbed power-law model with the photon 
index $\Gamma$ in the 1.8--1.9 range and the exponential cut-off energy 
$E_{\rm C}$ in the 150--400~keV range (e.g., \citealt{Gilli2007,Ueda2014}), 
accompanied by a Compton reflection with $R$ typically around 1. The 
Comptonization spectra can be approximated by the power-law cut-off model with 
$E_{\rm C}$ = 2--3 $kT_{\rm e}$, thus the plasma temperature in Seyferts should 
not usually exceed 200~keV. Nevertheless, due to a lack of detection of the 
emission above 200~keV for a large Seyfert sample the actual $E_{\rm C}$ remains 
quite uncertain and affects the predictions of the synthesis models 
\citep{Gilli2007}.  
 
The main goal of this work is a comprehensive spectral analysis of the 
high-quality medium/hard X-ray and soft $\gamma$-ray spectra of a large sample 
of Seyfert galaxies using a realistic, physical model. For this we use 
the hard X-ray/soft $\gamma$-ray data collected by the IBIS/ISGRI detector 
\citep{Lebrun2003} onboard the \integrale satellite over many years for 28 
Seyferts. These summed ISGRI spectra are accompanied by all
contemporary medium X-ray (3--18~keV) spectra from the \xmm, 
\suzakue and 
\xtee satellites, in order to model the spectral slope and absorption 
that cannot be studied with the ISGRI spectrum alone. The high-quality data from 
\xmm/EPIC pn and \suzaku/XIS allow a much better determination of the spectral 
parameters than it was possible with the X-ray detectors used with the OSSE data 
and those onboard of \sax. Compared to the analysis of \sax, 
\suzakue and \nust data from relatively short duration observations, our analysis 
should provide a better description of the mean spectral properties.  

\section{Data selection and reduction}
\label{dataset}

Our sample was selected from the second \integrale AGN catalog 
\citep{Beckmann2009}. We chose all the sources with a total ISGRI effective
exposure time larger than 500~ks\footnote{Corresponding to an observation with 
the source in the fully coded field of view.}. There are two more sources, 
namely NGC~4945 and Circinus galaxy, which fullfil the exposure criterion but 
we excluded them from our sample because their spectra are dominated by the 
reprocessed component.

\begin{table*}
\setlength{\tabcolsep}{1.3mm}
\centering
\caption{Sample of studied Seyfert galaxies and summed exposure time for each 
satellite. The objects in each group (Sy~1, Sy~1.5, Sy~2) are sorted in order of 
decreasing 20--100~keV flux. Central black hole masses $M_{\rm BH}$ are computed 
as a weighted mean of the values from the literature (see the text and Appendix 
\ref{appena}). The number of extracted spectra for each low-energy satellite is 
given in parentheses.}
\label{objects}
\footnotesize{
\begin{tabular}{lccrrrr}
\hline
Object          & Redshift & $\log(M_{\rm BH})$ [M$_{\sun}$]   & \integral    & \xmm     & \suzaku & \xte    \\
\multicolumn{3}{c}{} & \multicolumn{4}{c}{Exposure time [ks]} \\
\hline
\multicolumn{7}{c}{Type 1} \\
IC 4329A         & 0.0160   & 8.02$\pm$0.17                   & 1774     & 150 (2)  & 131 (5) & 382 (6)   \\
IGR J21247+5058  & 0.0200   & 7.80$\pm$0.50                   & 5656     & 40 (2)   & 85  (1) & ---       \\
GRS 1734-292     & 0.0214   & 8.94$\pm$0.30                   & 15214    & 18 (1)   & ---     & ---       \\
NGC 4593         & 0.0090   & 7.13$\pm$0.08                   & 2962     & 81 (2)   & 119 (1) & 902 (6)   \\
4U 0517+17       & 0.0179   & 7.18$\pm$0.14                   & 2399     & 62 (1)   & 46 (1)  & ---       \\
Akn 120          & 0.0337   & 8.18$\pm$0.06                   & 2209     & 112 (1)  & 101 (1) & 114 (1)   \\
ESO 141-55       & 0.0366   & 7.60$\pm$0.24                   & 979      & 213 (4)  & ---     & ---       \\
3C 111           & 0.0485   & 8.04$\pm$0.19                   & 2676     & 169 (2)  & 122 (1) & 676 (5)   \\
\hline
\multicolumn{7}{c}{Type 1.5} \\
NGC 4151         & 0.0033   & 7.43$\pm$0.11                   & 3236     & 354 (11) & 247 (3) & ---       \\
MCG+8-11-11      & 0.0205   & 8.02$\pm$0.04                   & 1404     & 38 (1)   & 99 (1)  & 17 (1)    \\
Mrk 509          & 0.0344   & 8.18$\pm$0.04                   & 1371     & 704 (6)  & 108 (4) & 439 (3)   \\
4U 1344-60       & 0.0130   & 7.67$\pm$0.36                   & 5781     & 37 (1)   & 94 (1)  & ---       \\
3C 120           & 0.0331   & 7.79$\pm$0.10                   & 1721     & 146 (2)  & 165 (4) & 1049 (5)  \\
NGC 6814         & 0.0052   & 7.14$\pm$0.11                   & 3664     & 32 (1)   & 42 (1)  & 19 (1)    \\
3C 390.3         & 0.0562   & 8.53$\pm$0.08                   & 1794     & 123 (2)  & 100 (1) & 247 (2)   \\
MCG-6-30-15      & 0.0079   & 6.86$\pm$0.18                   & 2174     & 462 (5)  & 384 (4) & 468 (5)   \\
\hline
\multicolumn{7}{c}{Type 2} \\
NGC 4388         & 0.0084   & 7.05$\pm$0.09                   & 4815     & 73 (2)   & 124 (1) & ---       \\
NGC 2110         & 0.0076   & 8.21$\pm$0.18                   & 1736     & 60 (1)   & 102 (1) & 180 (2)   \\
NGC 4507         & 0.0118   & 7.71$\pm$0.18                   & 1600     & 117 (5)  & 104 (1) & 137 (2)   \\
MCG-5-23-16      & 0.0085   & 7.45$\pm$0.20                   & 839      & 157 (2)  & 96 (1)  & 159 (2)   \\
NGC 5506         & 0.0062   & 7.31$\pm$0.18                   & 1479     & 264 (6)  & 159 (3) & 167 (2)   \\
Cygnus A         & 0.0561   & 9.41$\pm$0.10                   & 5614     & 41 (2)   & 45 (1)  & ---       \\
NGC 5252         & 0.0222   & 8.47$\pm$0.18                   & 3674     & 67 (1)   & 50 (1)  & ---       \\
ESO 103-35       & 0.0132   & 7.21$\pm$0.22                   & 1090     & 13 (1)   & 91 (1)  & 132 (1)   \\
NGC 788          & 0.0136   & 7.45$\pm$0.19                   & 2136     & 35 (1)   & 46 (1)  & ---       \\
NGC 6300         & 0.0037   & 6.63$\pm$0.22                   & 3580     & 53 (1)   & 83 (1)  & 26 (1)    \\
NGC 1142         & 0.0288   & 8.40$\pm$0.21                   & 1629     & 12 (1)   & 142 (2) & ---       \\
LEDA 170194      & 0.0367   & 8.58$\pm$0.50                   & 1135     & 17 (1)   & 130 (1) & ---       \\
\hline
\end{tabular}                                                                   
}                                                                               
\end{table*}     

Table \ref{objects} summarizes the basic physical properties of our sample. 
There are 8 type~1, 8 intermediate and 12 type~2 Seyfert nuclei, with at least 
one radio-loud object in each group. The BH mass estimates listed in Table 
\ref{objects} are the weighted 
mean of all estimates we found in the literature, as well as our estimates 
based on stellar velocity dispersion measurements, $\sigma_{\star}$. The data we 
used for the BH mass estimation, together with a brief description of the 
averaging procedure, are presented in Appendix \ref{appena}. The sources in our 
sample span a BH mass range of almost 3 orders of magnitude.

A summary of the observations we used for the spectral analysis is given in 
Table \ref{objects}. Table \ref{xspectra} in Appendix \ref{appenb} presents 
details of the selected spectral sets. We considered all the data collected by 
\integrale as of end of March 2010, except for 4U~0517+17, NGC~4151, 
MCG+8-11-11, NGC~4388, NGC~2110, NGC~5252 and ESO~103-35, because a large 
fraction of their data became public after that time (see Table \ref{xspectra}). 
Since the \integrale data were collected over many years, the summed ISGRI 
spectrum corresponds to the averaged emission of the objects in our sample. 

We selected data with off-axis angle $< 15\degr$\footnote{$<6\degr$ for 4U 
0517+17 due to its proximity to the Crab, which is contaminating the ISGRI
shadowgrams.} and we reduced them using the Offline 
Scientific Analysis (OSA) v.~9.0 provided by the INTEGRAL Science Data Centre
\citep{Courvoisier2003}. The standard spectral extraction software 
was used to extract the ISGRI spectrum for each \integrale pointing with the 
pipeline parameters set to the default values. We created an input source 
catalog including all sources detected within 15$\degr$ from each studied 
object. The ISGRI spectra were summed and the ancillary response files (ARFs) 
corresponding to them were generated using the OSA tool {\tt spe\_pick}. 
  
In the medium X-ray band we supplemented the ISGRI data with data from all 
available observations of the objects in our sample taken by \xmm, \suzakue 
and \xte. The \xmm/EPIC pn spectra were extracted using SAS 9.0.0 and the 
calibration files (CCF) as of 2009 October 21. The \xte/PCA and \suzaku/XIS spectra 
were extracted using HEASOFT 6.8.0 and the calibration files released before 
March 2010 through CALDB. For several objects we added to the spectral sets
the \xmme and \suzakue spectra collected in years 2010--2013. These new spectra
were extracted with SAS 13.5.0 and CCF of 2014 April 11 (\xmm) and HEASOFT 
6.16 and CALDB released on 2011 June 30 (\suzaku). In the case of the \suzaku/XIS 
we reprocessed the $3\times3$ and $5\times 5$ mode events and we merged them 
before the spectral extraction. The spectra of the front-illuminated CCDs (XIS0, 
XIS2 and XIS3) were added together, whereas the spectrum of the back-illuminated 
detector, XIS1, was used separately. We applied the standard method for 
extracting the XIS source spectra with the radius of the extraction region set 
to 250 pixels. However, for the background spectra we used several (two or 
three) smaller regions with a 125 pixels radius, as close to the source as 
possible. In this way we avoided the presence of instrumental background
lines seen in the background spectra when extracted for larger regions.
Depending which \xte/PCA PCU units were switched on during a given observation, 
we prepared the summed PCU 0+2 or PCU 2 spectra, merging all the data of a given 
observation ID.

\section{Spectral analysis}
\label{specana}

\subsection{Model definition}
\label{moddef}

Spectral fitting was performed with XSPEC v.~12.7.0 \citep{Arnaud1996}. Errors
are given for 90 per cent confidence level for a single parameter,
$\Delta\chi^2$ = $2.7$. 

For ISGRI we used the 18--200~keV range, except for 4U~0517+17, ESO~103-35 and 
LEDA~170194 where the spectra above 112~keV were affected by strong background 
fluctuations, as the observations of these sources were made predominantly at 
large off-axis angle. To limit the complexity of the fitted models, we did not 
consider X-ray data at energies below 3~keV. An additional rationale to not 
use the data below 3~keV is that, when we included them, the low-energy data 
dominated the fit in most sources. As a result, the ISGRI model fit worsened; 
large residuals appear above 20~keV and both the high-energy cut-off energy and 
the reflection fraction are not well determined. Consequently, the EPIC pn, the
XIS0-3 and the XIS1 spectra were used in the 3--10~keV, 3--10~keV and 3--9~keV 
bands, respectively. The only exceptions were ESO~141-55, Mrk~509, 4U~0517+17 
and LEDA~170194, for which we included the 1--3 keV data to the fit. The first 
two sources show a strong soft excess extending above 3~keV, where the data
were insufficient to constrain the absorption and excess parameters at the same 
time. 
The two other sources were 
among the weakest sources in our sample, in addition with only two low-energy 
observations for each. Thus, the extended low-energy spectra were needed to 
obtain a reasonable fit. The PCA PCU spectra were fitted in the 5--18~keV band. 
We excluded channels below 5~keV because we found that the standard correction 
of the xenon L-edge absorption at $\sim 4.1$~keV is not sufficient for the weak 
sources in the sample, and the residuals in this band were affecting the overall 
$\chi^2$ value. 

The model we adopted for spectral fitting is the following (in XSPEC 
terminology): {\tt CONSTANT*WABS*ZWABS*ZXIPCF*(COMPPS+ZGAUSS[+ZGAUSS])}. 
This is identical to the model used by \citet{Lubinski2010} in the case of 
NGC~4151, except that we have replaced the partial absorber model {\tt ZPCFABS} by 
{\tt ZXIPCF} which accounts for an ionized absorber partially covering the 
central source.

The component which accounts for the primary X--ray continuum emission is {\tt 
COMPPS}. This is a model for the estimation of thermal Comptonization spectra 
computed for different geometries using the exact numerical solution of the 
radiative transfer equation \citep{Poutanen1996}. The model also includes 
Compton reflection of the Comptonized emission from the cool 
medium \citep{Magdziarz1995}. We fitted the spectra allowing for four free 
parameters for this component: the temperature $kT_{\rm e}$ of the plasma cloud, 
the Compton parameter $y$, the reflection parameter $R$, and the normalization 
of the seed photons spectrum $K$. The Compton parameter can be used to determine 
the Thomson optical depth $\tau$ of the plasma cloud using the relation 
$\tau$ = $y m_{\rm e} c^2/(4kT_{\rm e}$, where $m_{\rm e}$ is 
the electron mass (e.g., \citealt{Rybicki1979}). The seed photons were assumed 
to have a multicolor disc blackbody distribution ({\tt  DISKBB} in XSPEC, 
\citealt{Mitsuda1984}) with the maximum temperature of $kT_{\rm bb}$ = 10~eV. 
As we have tested, fixing the maximum disc temperature at other value, e.g., 
100~eV, does not affect the fit results considerably.

The {\tt COMPPS} model offers several options for geometry of the plasma region 
but, as shown for NGC~4151 \citep{Lubinski2010}, none of them is statistically 
preferred even for the fit done to the highest-quality AGN spectra. Thus, we 
decided to use the simplest geometry option: a spherical case with an 
approximate treatment of radiative transfer using escape probability (denoted in 
{\tt COMPPS} by the geometry parameter = 0). The inclination angle of the 
reflector was set to $i$ = 30\degr, 45\degr{} and 60\degr{} for type 1, 1.5 and 
2 Seyferts, respectively. The reflector was assumed to be neutral, and 
relativistic broadening of the reflected spectra was ignored.  
Although the relativistic reflection from an ionized disc can be fitted to the 
spectra of particular AGN (e.g., \citealt{Reynolds2014}), these effects will be 
weak above 3~keV. We have tested this assumption for MCG--6-30-15, an 
archetypical AGN with relativistic effects, finding the relaxed disc ionization 
consitent with 0.

We added a redshifted Gaussian line to the primary emission component to model 
the iron K~$\rm\alpha$ emission from the host galaxy. For many objects we had 
also to include a second Gaussian corresponding to the iron K~$\rm\beta$ 
emission. The iron line parameters were fitted independently for each low-energy 
spectrum.

In addition to the continuum emission we also considered the case of absorption 
of the primary emission. The absorption consisted of three sub-components: 
Galactic (component {\tt WABS}) and redshifted neutral absorbers ({\tt ZWABS}),
and redshifted ionized, partially covering absorber ({\tt ZXIPCF}). The {\tt 
WABS} and {\tt ZWABS} models use the elemental abundances of \citet{Anders1982} 
and the photoelectric absorption cross-sections from \citet{Morrison1983}. The 
Galactic column density was set to a weighted mean given by the {\tt nH} HEASARC
tool and computed using the data from \citet{Kalberla2005}. 

The model {\tt ZXIPCF} uses a grid of XSTAR photoionized absorption models 
(calculated assuming a microturbulent velocity of 200~km s$^{-1}$) for the 
absorption. It assumes that the absorber covers some fraction $f^{\rm p}$ of the 
source, while the remaining fraction ($1-f^{\rm p}$) of the spectrum is seen 
directly \citep{Reeves2008}. Apart from $f^{p}$, the other parameters of this 
component are the column density of the ionized absorber $N_{H}^{p}$ and the 
logarithm of the ionization parameter $\xi^{p}$. The ionization parameter 
$\xi^{p}$ for \xte/PCA, the instrument with a low energy resolution, was set to 
the mean value obtained for the \xmm/EPIC pn and \suzaku/XIS spectra. After test 
fits we found that two other parameters of the complex absorber model, namely 
$N_{H}^{p}$ and $f^{p}$, have to be fixed to the mean obtained for the \xmme and 
\suzakue spectra, because the limited energy range of the \xtee spectra does not 
allow to constrain these parameters well.

For the pairs of \suzaku/XIS front and back-illuminated spectra we 
assumed the same spectral slope and absorption but their relative normalizations 
and parameters of iron lines were allowed to be different, to account for the 
difference in their absolute calibration and energy resolution. After some initial 
tests we applied a 3 per cent systematic error to the \suzaku/XIS spectra, all other 
spectra were fitted with the statistical errors only.

For two sources with the \xmme and \suzakue spectra fitted in the 1--10~keV 
band, namely ESO~141-55 and Mrk~509, we had to add several additional 
components. In the case of ESO~141-55 these were two absorption edges around 2.1 
and 2.6~keV (XSPEC model {\tt ZEDGE}), bremsstrahlung emission model 
({\tt ZBREMSS}) for the soft excess and an additional emission line at 
$\approx 2.5$~keV. Only one absorption edge at $\approx$1.8~keV was 
needed for the Mrk~509 spectra, together with the bremsstrahlung component 
modelling the soft excess. 

\begin{figure}
\includegraphics[width=82mm]{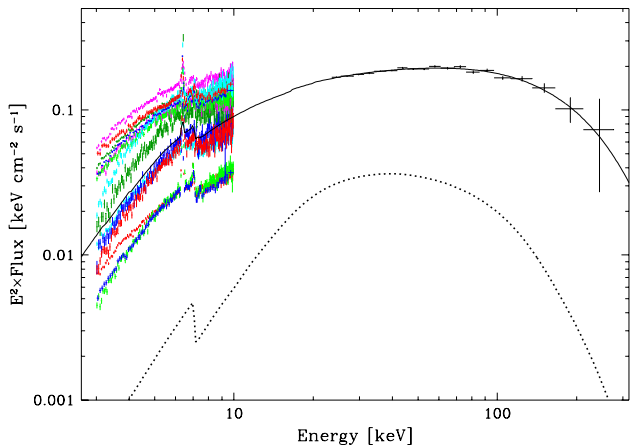}
\caption{Spectral modelling (option D) applied to the spectra of NGC~4151. The 
unfolded ISGRI spectrum is shown in black and spectra taken with \xmme and 
\suzakue are shown in colour. The solid line shows an absorbed continuum model 
(with reflection) fitted to the ISGRI spectrum, the dotted line shows the 
reflection component. The models fitted to the low-energy spectra are not 
shown.}
\label{sp4151}
\end{figure}

\subsection{Model options}
\label{modopt}

For each source the ISGRI spectrum and all low-energy X--ray 
spectra were fitted simultaneously with the model described 
above. The ISGRI spectra are collected over a long time, while the low-energy 
observations are usually short, especially in the case of \xmme and \suzaku. 
These observations should provide information about ``short" term spectral 
variations for each object. Regarding these variations, we considered the 
possibility of spectral slope and/or warm absorber variations. We did not 
consider variations of $kT_{\rm e}$ and $R$, as these parameters cannot be determined 
from the model fits to the individual low-energy spectrum separately, due to 
their limited energy coverage. For that reason, during the spectral fits we kept 
$kT_{\rm e}$ and $R$ tied to the same value for all spectra.

Table \ref{modoptions} summarizes the different variability options 
that we considered for the spectral slope and absorption. First we fitted all 
the data sets for each object assuming no spectral variations, i.e., all the 
model parameters were kept tied to the same value, for all spectra, apart from 
the model normalization which was always allowed to vary (model A). Then we 
assumed that only the spectral slope, 
characterized by $y$, was variable (model B) or that only the absorption was 
variable (model C). We also considered the possibility that both the spectral 
slope and the absorption was variable (model D). In addition to the model 
options presented already, for each object we made a fit 
alternative to model B, denoted as Ba, with the partially covering absorption 
component removed. This test allowed us to judge how the choice of the 
absorption model, done in various analyses, affects the results.

During all model fits the ISGRI normalization was fixed at 1. 
Depending on the model version we were fitting to the data, the model parameters 
`checked' in Table \ref{modoptions} were left as free variables for all 
low-energy spectra. The respective model parameter for the ISGRI spectrum was 
kept fixed to the arithmetic mean of all the best-fitting parameter values 
determined from each one of the low-energy spectra. The other continuum model 
parameters were kept tied to the same value during the fit, for all spectra.
In summary, our approach assumes that the ISGRI spectrum represents the average 
spectral shape for all objects in our sample, while the low-energy spectra can 
be used to detect temporal deviations of the spectral slope (models B, Ba, D) 
and warm absorption (models C, D) from the respective average model parameter 
values.

\begin{table}
\setlength{\tabcolsep}{1.3mm}
\centering
\caption{Summary of the spectral model options applied in the fit. The 
parameters allowed to be different for each low-energy spectrum are checkmarked, 
whereas the parameters not present in the model Ba are marked with em-dash.} 
\label{modoptions}
\footnotesize{
\begin{tabular}{lcccccc}
\hline
Model option & \multicolumn{6}{c}{Parameter} \\
       & $n_{I}$    & $y$   & $N_{H}$ & $N_{Hp}$ & $f_{p}$ & $\log \xi_{p}$ \\
\hline
A   & $\checkmark$ &              &              &              &              &              \\
B   & $\checkmark$ & $\checkmark$ &              &              &              &              \\
Ba  & $\checkmark$ & $\checkmark$ &              & ---          & ---          & ---          \\
C   & $\checkmark$ &              & $\checkmark$ & $\checkmark$ & $\checkmark$ & $\checkmark$ \\
D   & $\checkmark$ & $\checkmark$ & $\checkmark$ & $\checkmark$ & $\checkmark$ & $\checkmark$ \\
\hline
\end{tabular}                                                                   
}                                                                               
\end{table}         

\begin{figure}
\includegraphics[width=82mm]{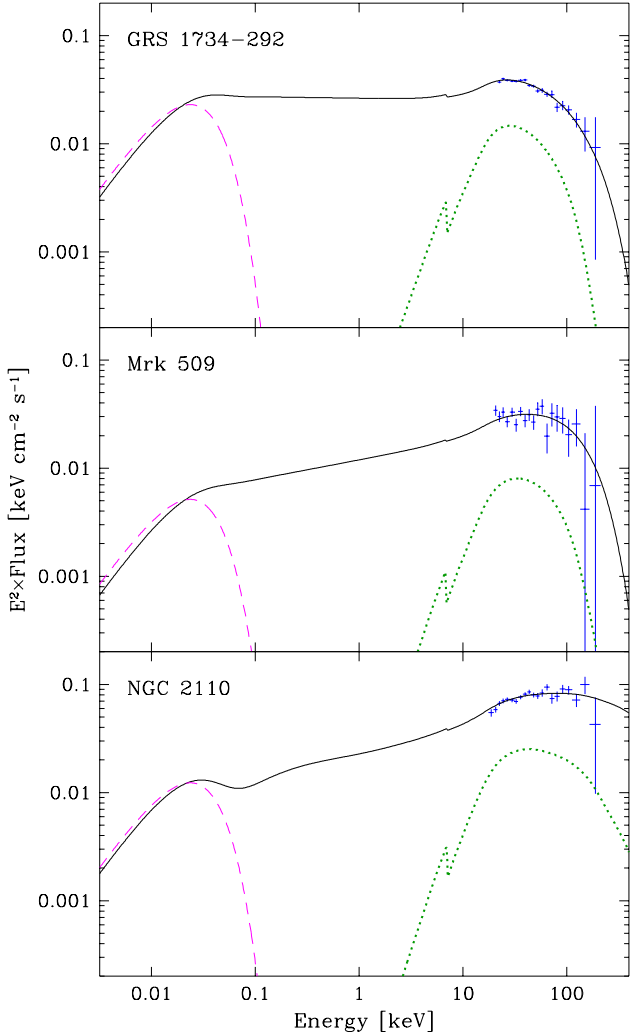}
\caption{Unfolded ISGRI (blue) spectra of GRS~1734-292, Mrk~509 and NGC~2110. 
Solid line - an unabsorbed Comptonization plus reflection model, dashed 
line - multi-colour black body model of seed photons, dotted line - the 
reflection component.}
\label{specs}
\end{figure}

An illustration of our spectral modelling is presented in Figure \ref{sp4151}, 
showing an entire spectral set used for the NGC~4151 study together with the 
total and reflection models fitted to the average spectrum. The data from \xmme 
and \suzakue show a range of variability observed for the low-energy spectra.
Figure \ref{specs} presents the examples of the best model fitted to the 
spectral sets of GRS~1734-292, Mrk~509 and NGC~2110. In this figure we show only 
the mean model corresponding to the ISGRI spectrum and the ISGRI data. Note the 
high quality of the GRS~1734-292 spectrum, thanks to the 15~Ms exposure time.
This spectrum also shows that the ISGRI spectra extracted with OSA~9 are not 
affected by strong systematic features: data points agree with the model within 
the error bars. The rest of the ISGRI spectra is presented in the Appendix
\ref{appenc}. 

During the spectral fitting our major concern was a careful test if the global
$\chi^{2}$ minimum was found for a given model option and data set. Modelling
with XSPEC or any other spectral software cannot be fully automatized,
especially when the model is complex and the spectra have a large number of 
channels. Therefore we performed initial tests looking for the best-fitting 
value of $K$, $kT_{\rm e}$, $y$ and $R$ parameters, scanning their values in a 
wide range using the XSPEC {\tt steppar} command. Then we computed in the same 
way the uncertainties for all other free parameters. Finally, the fit was 
checked running two-dimensional scan for several pairs of parameters: 
$kT_{\rm e}-R$ for model options B and D and $kT_{\rm e}-R$, $kT_{\rm e}-y$ and 
$y-R$ for model option C. Whenever we found a new $\chi^{2}$ minimum during one 
of these phases of analysis, all the phases were repeated until there was no 
more fit improvement.

\subsection{Method limitations}

The most correct way of determining the average spectral properties of any class 
of objects is to compute a weighted mean for the parameters determined for 
possibly short observations sampling frequently a possibly long period. However, 
the number of emitted photons and instrumental background change rapidly with 
increasing energy in the X-ray and soft $\Gamma$-ray domain, reducing strongly 
data quality in the high-energy band. For this reason the parameters driven 
mainly by that band cannot be well determined with the short exposure time 
spectra. Our strategy of using the summed ISGRI spectra allows us to find 
accurate mean values of $kT_{\rm e}$ and $R$ at the cost of a model where some 
parameters are set to the mean and some are fitted separately to the low-energy
spectra. Thus, the determined values of $kT_{\rm e}$ and especially $R$ can be 
biased in situation when both these quantities vary between low-energy 
observations. A scale of that bias can be estimated with the fit done to the 
spectra simulated with the models assuming various plausible variability 
patterns. Since finding a global $\chi^{2}$ minimum for a single case is work 
intensive and time consuming as described above, a massive fitting of simulated 
spectra is practically impossible. Instead, we have tried several simpler ways 
to check the reliability of our results.

One of these tests is the application of various model options presented above. 
For a vast majority of studied Seyferts the results of models B, C and D 
assuming variable Compton parameter and/or absorption appeared quite similar 
(see Table \ref{averages}). On the other hand, we included in our sample the 
X-ray brightest Seyfert, namely NGC~4151, for which an analysis of the 
flux-resolved spectral sets, including contemporary low- and high-energy spectra, 
was done earlier with a similar complex model \citep{Lubinski2010}. The mean 
$kT_{\rm e}$, $y$ and $R$ values found here (see Table \ref{mainres}) are very 
similar to those determined by \citet{Lubinski2010} for the bright state of 
NGC~4151. This is expected as the summed spectrum will be always dominated by 
the bright state, which in addition prevails in the \integrale observations.
Thus, at least for NGC~4151 our `average' approach produces reasonable results.
In addition, for this bright object we did a special fit with $R$ fitted 
separately for each low-energy spectrum. Although the fitted $R$ values vary 
between 0 and 1.4, their mean equal to 0.34$_{-0.07}^{+0.12}$ is almost the same 
as that fitted with model D. The $kT_{\rm e}$ and $y$ values do not change as well.     
Therefore, a constant $R$ approximation appears to be justified. 

\section{Results}
\label{results}

\subsection{The best-fitting model selection}
\label{selmod}

The best-fitting results for all model options should be compared in a 
quantitative way to choose the final result for a given object. A commonly 
applied criterion to convert the $\chi^{2}$ values obtained for two models into 
their relative probability measure is the $F$-test. Although this test is 
justified in many situations, its assumptions allowing for a computing of the 
probability $P_{\rm F}$ based on the $F$-distribution are not always met. When 
the models are not nested or when the interesting parameter is set to its 
boundary limit in the simpler model, the true distribution can be far from the 
reference test distribution \citep{Protassov2002}. The model options used in our 
analysis violate both these conditions. Despite the fact that we use the same 
general XSPEC model form for all alternatives, any operation tying one parameter 
to the other, e.g., when the absorption or Compton $y$ parameters are computed 
as an average for the ISGRI spectrum, breaks the nesting condition. The second 
assumption is not met when we fix, for example, the $N_{\rm H}^{\rm p}$ and 
$f^{\rm p}$ parameters of the complex absorber component to zero, fitting the 
simple absorption model option Ba. 

Model selection is still an open issue of the contemporary statistics, with 
various selection methods showing their pros and cons (e.g., 
\citealt{Wit2012}). One of the most general and robust tests is the Bayesian 
model comparison, either in the form of exact Bayes factors or reduced to the 
Bayesian information criterion (BIC). Since the Bayesian analysis 
is computationally demanding, we decided to 
use an easy to compute and robust test in the form of the Akaike information 
criterion (AIC, \citealt{Akaike1973}), for which the nesting condition is not
required. To take into account a bias introduced by a finite size of the sample, 
we applied a corrected AIC test \citep{Sugiura1978}, which for the
$\chi^{2}$-statistic used for a spectral fit has the form
\begin{equation}
AIC_{\rm C} = 2k-2C_{\rm L}+\chi^{2} + \frac{2k(k+1)}{n-k-1},
\end{equation}
where $C_{\rm L}$ is the likelihood of the true hypothetical model and does not 
depend on the data or tested models, $k$ is the number of free parameters of a 
given model and $n$ is the total number of the channels in the fitted spectra. 
Since the models are compared through computing a difference of their 
$AIC_{\rm C}$ values, the constant $C_{\rm L}$ cancels out. The relative 
likelihood $p_{\rm A}$ that the more complex model $b$ is better than its 
simpler alternative $s$ is given by the exponential $p_{\rm A}$ = 
$\exp[0.5(AIC_{\rm C}^{\rm b}-AIC_{\rm C}^{\rm s})]$. In our case, the reference 
(most complex) model was the model D, to which all other simpler model options 
were compared. 

The Akaike criterion strongly favours the models with fewer parameters due to 
the terms with the free parameter number $k$. In the case of our analysis 
models B with a variable spectral slope are favoured over the models C with a 
variable absorption, because the latter need three free parameters more for a 
single low-energy spectrum. The X-ray absorption varying on time scales of days, 
months or years is a well established fact for the Seyfert galaxies, whenever 
a high-quality spectral sets were available (e.g., 
\citealt{Puccetti2007,Turner2009}). Thus, the penalty of more complex models 
introduced with the Akaike test appears too strong. We observe typically much 
larger uncertainties of the $kT_{\rm e}$, $y$ and $R$ values for model B than 
for models C and D. This indicates that the variability of the $y$ parameter is 
not sufficient to explain the observed spectral variability. For this reason we 
decided to add an additional model selection criterion, requiring the lowest 
$\chi^{2}$ for the ISGRI spectrum alone. This addition is strongly required as 
the two basic parameters we seek, namely $kT_{\rm e}$ and $R$, are determined 
mainly by the ISGRI spectrum. Consequently, the final model selection criterion 
was: a model with the highest $p_{\rm A}$ value that is giving the lowest 
$\chi^{2}$ for the ISGRI spectrum. 

\subsection{The best-fitting results}
\label{fitres}

\begin{table*}
\setlength{\tabcolsep}{3pt}
\centering
\caption{Main results of the spectral fitting. Comptonization model parameters: 
$kT_{\rm e}$ - electron plasma temperature in keV; $y$ - Compton parameter; 
$\tau$ - plasma optical depth; $R$ - relative amplitude of the reflection 
component; $K$ - absolute normalization of the seed photons spectrum in 
$10^{8}$. Other {\tt COMPPS} parameters were fixed: the temperature of the seed 
photons (a multi-colour black body) $T_{\rm bb}$ = 10 eV, the inclination angle 
$i$ = 30\degr (Sy 1), 45\degr (Sy 1.5), 60\degr (Sy 2), the disc temperature 
$temp$ = 10$^{5}$ K. Other parameters: $N_{\rm H}$, $N_{\rm H}^{\rm p}$ - column 
density of neutral and ionized, partially covering absorbers, respectively, in 
$10^{22}$ cm$^{-2}$, $f^{\rm p}$ and $\xi^{\rm p}$ (in erg cm s$^{-1}$) - 
covering fraction and ionization parameter, respectively, of partial, ionized 
absorber. The photon index $\Gamma$ and the plasma optical depth $\tau_{\rm MC}$ 
were computed in a way described in the text. For model options see Table 
\ref{modoptions}.}
\small
\label{mainres}
\begin{tabular}{lcccccccccccc}

\hline

Object (model) & $kT_{\rm e}$ & $R$ &  $K$ & $y$ & $\Gamma$ & $\tau$ & $\tau_{\rm MC}$ & 
$N_{\rm H}$ & $N_{\rm H}^{\rm p}$ & $f^{\rm p}$ & $\log \xi^{\rm p}$ & $\chi ^{2}_{\rm red}$ \\

\hline
IC 4329A (D) & 40$_{-5}^{+7}$ & 0.12$_{-0.04}^{+0.04}$ 
& 3.75$_{-0.14}^{+0.56}$ & 0.95$_{-0.02}^{+0.02}$ & 1.87$_{-0.02}^{+0.02}$ 
& 3.0$_{-0.6}^{+0.4}$ & 2.25 & 0.43$_{-0.17}^{+0.17}$ & 103$_{-10}^{+14}$ 
& 0.23$_{-0.02}^{+0.04}$ & 1.62$_{-0.19}^{+0.15}$ & 0.90 \\[2pt]

IGR J21247+5058 (C) & 35$_{-6}^{+10}$ & 0.11$_{-0.11}^{+0.25}$ 
& 0.93$_{-0.17}^{+0.27}$ & 1.14$_{-0.06}^{+0.05}$ & 1.74$_{-0.03}^{+0.04}$ 
& 4.2$_{-1.2}^{+0.8}$ & 2.9 & 2.4$_{-0.8}^{+0.4}$ & 102$_{-27}^{+50}$ 
& 0.35$_{-0.09}^{+0.16}$ & 0.7$_{-1.3}^{+0.6}$ & 0.95 \\[2pt]

GRS 1734-292 (A) & 49$_{-16}^{+20}$ & 0.66$_{-0.44}^{+0.39}$ 
& 3.42$_{-0.96}^{+2.23}$ & 0.78$_{-0.09}^{+0.07}$ & 2.01$_{-0.07}^{+0.12}$ 
& 2.0$_{-0.9}^{+0.7}$ & 1.73 & 2.8$_{-1.4}^{+0.4}$ & 31$_{-21}^{+391}$ & 
0.27$_{-0.09}^{+0.13}$ & -0.6$_{-2.4}^{+1.1}$ & 1.08 \\[2pt]

NGC 4593 (B) & 27$_{-4}^{+6}$ & 0.0$_{-0.0}^{+0.07}$ 
& 0.61$_{-0.07}^{+0.13}$ & 1.09$_{-0.02}^{+0.02}$ & 1.77$_{-0.02}^{+0.01}$ 
& 5.2$_{-1.1}^{+0.8}$ & 3.45 & 0.0$_{-0.0}^{+0.24}$ & 143$_{-21}^{+28}$ 
& 0.22$_{-0.03}^{+0.02}$ & 2.84$_{-0.07}^{+0.07}$ & 1.00 \\[2pt]

4U 0517+17 (D) & 32$_{-8}^{+16}$ & 0.40$_{-0.35}^{+0.30}$ & 0.42$_{-0.07}^{+0.08}$ 
& 1.09$_{-0.05}^{+0.05}$ & 1.77$_{-0.03}^{+0.03}$ & 4.35$_{-2.19}^{+1.11}$ & 3.0 & 
0.06$_{-0.02}^{+0.01}$ & 8.8$_{-4.0}^{+2.5}$ & 0.61$_{-0.25}^{+0.05}$ 
& 1.46$_{-0.48}^{+0.46}$ & 0.90 \\[2pt]

Akn 120 (D) & 179$_{-28}^{+23}$ & 0.49$_{-0.19}^{+0.16}$ 
& 8.39$_{-1.53}^{+1.13}$ & 0.44$_{-0.03}^{+0.04}$ & 2.13$_{-0.07}^{+0.02}$ 
& 0.31$_{-0.05}^{+0.06}$ & 0.24 & 0.05$_{-0.05}^{+0.40}$ & 44$_{-7}^{+10}$ 
& 0.30$_{-0.04}^{+0.04}$ & -0.86$_{-0.30}^{+0.92}$ & 0.96 \\[2pt]

ESO 141-55 (B)
& 138$_{-33}^{+18}$ & 0.77$_{-0.07}^{+0.14}$ & 2.17$_{-0.17}^{+0.20}$ 
& 0.64$_{-0.04}^{+0.02}$ & 2.02$_{-0.04}^{+0.09}$ & 0.59$_{-0.09}^{+0.13}$ & 0.47
& 0.0$_{-0.0}^{+0.012}$ & 0.31$_{-0.05}^{+0.07}$ & 0.37$_{-0.03}^{+0.03}$  
& -0.27$_{-0.04}^{+0.03}$ & 1.32 \\[2pt]

3C 111 (B) & 113$_{-68}^{+107}$ & 0.26$_{-0.07}^{+0.08}$ 
& 1.18$_{-0.16}^{+1.88}$ & 0.94$_{-0.08}^{+0.06}$ & 1.84$_{-0.10}^{+0.10}$ 
& 1.1$_{-1.0}^{+0.7}$ & 0.85 & 0$_{-0.0}^{+0.8}$ & 4.7$_{-0.8}^{+14.0}$ & 
0.25$_{-0.12}^{+0.04}$ & -3$_{-0.0}^{+3.6}$ & 1.05 \\

\hline

NGC 4151 (D) & 53$_{-3}^{+7}$ & 0.33$_{-0.05}^{+0.08}$ 
& 3.41$_{-0.36}^{+0.33}$ & 1.11$_{-0.04}^{+0.04}$ & 1.76$_{-0.02}^{+0.03}$ 
& 2.68$_{-0.37}^{+0.18}$ & 2.04 & 3.7$_{-0.4}^{+0.9}$ & 21.3$_{-1.4}^{+1.6}$ 
& 0.75$_{-0.05}^{+0.02}$ & 1.06$_{-0.22}^{+0.06}$ & 1.12 \\[2pt]

MCG+8-11-11 (D) & 44$_{-12}^{+23}$ & 0.0$_{-0.0}^{+0.12}$ & 4.3$_{-1.5}^{+2.0}$ 
& 0.85$_{-0.08}^{+0.05}$ & 1.95$_{-0.05}^{+0.09}$ & 2.5$_{-1.3}^{+0.7}$ & 1.85
& 1.9$_{-1.2}^{+2.6}$ & 100$_{-9}^{+49}$ & 0.43$_{-0.09}^{+0.10}$ & 0.8$_{-0.9}^{+0.3}$ 
& 0.83 \\[2pt]

Mrk 509 (C) & 43$_{-12}^{+27}$ & 0.40$_{-0.04}^{+0.03}$ 
& 0.76$_{-0.04}^{+0.04}$ & 1.03$_{-0.02}^{+0.02}$ & 1.81$_{-0.02}^{+0.02}$ 
& 3.1$_{-2.0}^{+0.9}$ & 2.20 & 0.04$_{-0.04}^{+0.10}$ & 1.48$_{-0.18}^{+0.11}$ 
& 0.43$_{-0.07}^{+0.05}$ & -2.43$_{-0.25}^{+0.39}$ & 1.06 \\[2pt]

4U 1344-60 (C) & 30$_{-8}^{+19}$ & 0.55$_{-0.55}^{+0.73}$ & 0.62$_{-0.19}^{+0.28}$ 
& 1.05$_{-0.08}^{+0.10}$ & 1.79$_{-0.06}^{+0.06}$ & 4.5$_{-2.9}^{+1.3}$ 
& 3.1 & 1.9$_{-0.4}^{+0.5}$ & 182$_{-57}^{+99}$ & 0.53$_{-0.13}^{+0.11}$ 
& 2.80$_{-0.21}^{+0.10}$ & 0.94 \\[2pt]

3C 120 (D) & 176$_{-23}^{+24}$ & 0.27$_{-0.06}^{+0.07}$ 
& 8.2$_{-1.6}^{+1.8}$ & 0.50$_{-0.03}^{+0.03}$ & 2.08$_{-0.06}^{+0.07}$ 
& 0.36$_{-0.05}^{+0.05}$ & 0.3 & 0.52$_{-0.11}^{+0.19}$ & 36.7$_{-3.4}^{+4.2}$ 
& 0.37$_{-0.02}^{+0.02}$ & -1.34$_{-0.08}^{+0.35}$ & 0.96 \\[2pt]

NGC 6814 (C) & 46$_{-14}^{+69}$ & 0.0$_{-0.0}^{+0.22}$ 
& 0.42$_{-0.13}^{+0.48}$ & 1.14$_{-0.16}^{+0.08}$ & 1.74$_{-0.05}^{+0.08}$ 
& 3.2$_{-3.2}^{+1.0}$ & 2.35 & 0.1$_{-0.1}^{+1.0}$ & 101$_{-56}^{+49}$ 
& 0.37$_{-0.13}^{+0.10}$ & 1.86$_{-0.48}^{+0.28}$ & 1.11 \\[2pt]

3C 390.3 (B) & 37$_{-11}^{+44}$ & 0.0$_{-0.0}^{+0.15}$ & 0.66$_{-0.21}^{+0.24}$ 
& 1.06$_{-0.03}^{+0.03}$ & 1.79$_{-0.02}^{+0.02}$ & 3.7$_{-3.7}^{+1.1}$ 
& 2.55 & 0.17$_{-0.17}^{+0.18}$ & 81$_{-33}^{+18}$ & 0.21$_{-0.08}^{+0.02}$ 
& 1.90$_{-0.45}^{+0.13}$ & 0.96 \\[2pt]

MCG-6-30-15 (D) & 31$_{-5}^{+9}$ & 1.77$_{-0.14}^{+0.12}$ 
& 6.25$_{-0.18}^{+0.48}$ & 0.60$_{-0.01}^{+0.01}$ & 2.26$_{-0.03}^{+0.03}$ 
& 2.5$_{-0.7}^{+0.4}$ & 1.88 & 2.35$_{-0.24}^{+0.21}$ & 151$_{-35}^{+46}$ 
& 0.43$_{-0.09}^{+0.08}$ & 4.37$_{-0.11}^{+0.11}$ & 1.02 \\

\hline
NGC 4388 (D) & 53$_{-9}^{+17}$ & 0.07$_{-0.07}^{+0.17}$ 
& 1.83$_{-0.26}^{+0.40}$ & 1.11$_{-0.11}^{+0.11}$ & 1.76$_{-0.06}^{+0.07}$ 
& 2.7$_{-0.9}^{+0.6}$ & 1.92 & 20.9$_{-6.4}^{+1.4}$ & 57$_{-21}^{+48}$ & 0.63$_{-0.06}^{+0.08}$ 
& 1.79$_{-0.22}^{+0.13}$ & 1.10 \\[2pt]

NGC 2110 (D) & 230$_{-57}^{+51}$ & 0.63$_{-0.16}^{+0.09}$ 
& 1.83$_{-0.33}^{+0.62}$ & 0.94$_{-0.09}^{+0.10}$ & 1.75$_{-0.06}^{+0.09}$ & 0.52$_{-0.13}^{+0.14}$
& 0.39 & 1.0$_{-0.4}^{+1.3}$ & 5.9$_{-3.3}^{+3.7}$ & 0.54$_{-0.17}^{+0.17}$ & -0.47$_{-0.16}^{+0.51}$ 
& 0.89 \\[2pt]

NGC 4507 (D) & 30$_{-4}^{+5}$ & 0.0$_{-0.0}^{+0.09}$ 
& 0.84$_{-0.14}^{+0.29}$ & 1.19$_{-0.13}^{+0.11}$ & 1.71$_{-0.06}^{+0.08}$ 
& 5.1$_{-1.0}^{+0.8}$ & 3.4 & 7.1$_{-1.8}^{+1.8}$ & 54.6$_{-2.5}^{+3.5}$ 
& 0.95$_{-0.03}^{+0.03}$ & 1.13$_{-0.10}^{+0.09}$ & 1.14 \\[2pt]

MCG-5-23-16 (D) & 97$_{-38}^{+38}$ & 0.69$_{-0.22}^{+0.28}$ 
& 8.3$_{-2.9}^{+3.7}$ & 0.69$_{-0.08}^{+0.09}$ & 2.04$_{-0.12}^{+0.15}$ 
& 0.91$_{-0.37}^{+0.38}$ & 0.75 & 1.87$_{-0.58}^{+0.54}$ & 32.5$_{-5.8}^{+6.7}$ 
& 0.38$_{-0.04}^{+0.04}$ & 0.49$_{-0.51}^{+0.28}$ & 0.98 \\[2pt]

NGC 5506 (D) & 28$_{-3}^{+5}$ & 0.31$_{-0.08}^{+0.06}$ 
& 5.27$_{-0.46}^{+0.53}$ & 0.84$_{-0.02}^{+0.02}$ & 1.97$_{-0.02}^{+0.02}$ 
& 3.8$_{-0.7}^{+0.4}$ & 2.7 & 4.0$_{-0.2}^{+0.3}$ & 218$_{-23}^{+26}$ 
& 0.34$_{-0.05}^{+0.07}$ & 2.69$_{-0.09}^{+0.04}$ & 0.99 \\[2pt]

Cygnus A (B) & 185$_{-11}^{+10}$ & 0.32$_{-0.06}^{+0.05}$ 
& 3.46$_{-0.13}^{+0.18}$ & 0.69$_{-0.02}^{+0.02}$ & 1.92$_{-0.02}^{+0.03}$ 
& 0.48$_{-0.03}^{+0.03}$ & 0.37 & 0.0$_{-0.0}^{+0.4}$ & 11.1$_{-1.4}^{+2.4}$ 
& 0.63$_{-0.04}^{+0.02}$ & 0.35$_{-0.10}^{+0.25}$ & 1.09 \\[2pt]

NGC 5252 (A) & 84$_{-36}^{+491}$ & 0.0$_{-0.0}^{+0.63}$ 
& 0.78$_{-0.36}^{+2.52}$ & 1.12$_{-0.64}^{+0.12}$ & 1.75$_{-0.07}^{+0.67}$ 
& 1.70$_{-1.70}^{+0.76}$ & 1.4 & 0.0$_{-0.0}^{+6.6}$ & 5.8$_{-2.5}^{+22.3}$ 
& 0.98$_{-0.88}^{+0.02}$ & 0.3$_{-3.3}^{+1.5}$ & 0.92 \\[2pt]

ESO 103-35 (C) & 43$_{-14}^{+46}$ & 0.88$_{-0.30}^{+0.62}$ 
& 4.24$_{-1.33}^{+1.92}$ & 0.74$_{-0.08}^{+0.07}$ & 2.06$_{-0.11}^{+0.12}$ 
& 2.20$_{-2.20}^{+0.75}$ & 1.65 & 21.1$_{-0.5}^{+0.5}$ & 190$_{-38}^{+94}$ 
& 0.40$_{-0.13}^{+0.17}$ & 2.77$_{-0.31}^{+0.72}$ & 0.89 \\[2pt]

NGC 788 (D) & 54$_{-22}^{+44}$ & 0.08$_{-0.08}^{+1.20}$ 
& 0.7$_{-0.5}^{+1.2}$ & 1.06$_{-0.16}^{+0.21}$ & 1.79$_{-0.12}^{+0.12}$ & 2.5$_{-2.1}^{+1.2}$ 
& 1.95 & 13.7$_{-9.7}^{+9.9}$ & 101$_{-29}^{+14}$ & 0.99$_{-0.02}^{+0.01}$ 
& 1.44$_{-0.10}^{+0.13}$ & 0.98 \\[2pt]

NGC 6300 (C) & 62$_{-30}^{+245}$ & 2.32$_{-0.82}^{+0.97}$ 
& 0.61$_{-0.27}^{+2.54}$ & 0.91$_{-0.50}^{+0.11}$ & 1.89$_{-0.20}^{+0.76}$ 
& 1.9$_{-1.9}^{+1.0}$ & 1.45 & 24.2$_{-2.4}^{+2.4}$ & 178$_{-124}^{+87}$ 
& 0.69$_{-0.19}^{+0.15}$ & 4.25$_{-0.27}^{+0.41}$ & 0.97 \\[2pt]

NGC 1142 (D) & 359$_{-117}^{+133}$ & 1.3$_{-1.3}^{+2.8}$ 
& 0.80$_{-0.50}^{+1.04}$ & 0.69$_{-0.21}^{+0.32}$ & 1.81$_{-0.09}^{+0.19}$ & 0.25$_{-0.12}^{+0.14}$
& 0.19 & 6.6$_{-6.1}^{+8.5}$ & 72$_{-13}^{+39}$ & 0.98$_{-0.02}^{+0.01}$ & 1.26$_{-0.20}^{+0.26}$ 
& 0.86 \\[2pt]

LEDA 170194 (C) & 26$_{-10}^{+229}$ & 0.0$_{-0.0}^{+0.73}$ 
& 0.27$_{-0.09}^{+0.27}$ & 1.15$_{-0.67}^{+0.11}$ & 1.73$_{-0.10}^{+0.80}$ 
& 5.7$_{-5.7}^{+2.2}$ & 3.65 & 0.5$_{-0.4}^{+0.5}$ & 5.7$_{-0.5}^{+0.6}$ 
& 0.96$_{-0.05}^{+0.02}$ & 0.13$_{-1.00}^{+0.13}$ & 0.80 \\

\hline

\end{tabular}
\end{table*}

Table \ref{mainres} presents the summary of the spectral fitting results,
showing only the parameters of the best mean continuum model selected for each
object. The most complex model option D, assuming variable both spectral slope 
and absorption, almost always resulted in the smallest best-fitting $\chi^{2}$ 
value. The only exception was GRS~1734-292 with a single low-energy spectrum,
thus allowing to fit only model option A. Only half of the D option cases 
passed the Akaike criterion test preferring a simpler option A, B or C. A 
secondary criterion selecting the model option with the smallest $\chi^{2}$ 
value for the ISGRI spectrum had to be applied only three times, changing the 
choice from model C to D for 4U~0517+17 ($\Delta \chi^{2}/\chi^{2}$ = 
0.3/16.0) and from model B to D for MCG+8-11-11 ($\Delta \chi^{2}/\chi^{2}$ =
1.0/36.0) and for 3C~120 ($\Delta \chi^{2}/\chi^{2}$ = 1.5/20.9). Model B with 
the varying low-energy spectral slope appeared the best for 5 objects, mainly 
those less absorbed. For 7 objects the best choice was option C with variable 
absorption. The number of high-quality low-energy spectra strongly determines 
the best model form: for all cases where there were at least 4 spectra taken 
with either \xmme and/or \suzakue the best model is model D. We conclude that, 
apart from significant flux variations, Seyferts usually show significant 
spectral slope and/or warm absorbing variations as well.

Besides the parameters fitted directly to the spectra with a model defined in 
Sec. \ref{moddef} we have computed three other parameters based on the fit 
results. The first is the photon index $\Gamma$ used commonly in 
phenomenological power-law models approximating the continuum of the X-ray 
spectra of accreting systems. To compute a value of $\Gamma$ corresponding to 
our results, we simulated a large number of \xmme spectra with the {\tt 
COMPPS} model and a dense grid of ($kT_{\rm e},y$) values. These spectra were 
fitted in the 2--10~keV band with a power-law model, providing an array of 
$\Gamma$ values. Then the $\Gamma$ value for each ($kT_{\rm e},y$) pair, 
determined with the {\tt COMPPS} model for our AGN sample, was computed with a 
two-dimensional interpolation. The Thomson optical depth $\tau$ of the plasma 
was computed using the best-fitting values of the $kT_{\rm e}$ and $y$ 
parameters (see Sec. \ref{moddef}). Since the number of Compton scatterings 
considered in the {\tt COMPPS} model is limited, the $\tau$ value computed this 
way can be incorrect. The authors of the code warn of situations where $\tau$ 
$>$ 3 and this limit is exceeded for many spectral sets fitted here. Therefore 
all $\tau$ values were checked with a Monte Carlo code simulating the 
Comptonization process \citep{Gierlinski2000}, based on a model of 
\citet{Gorecki1984}. The fitted $kT_{\rm e}$ was assumed and then the 
simulated $\tau_{\rm MC}$ was adjusted to obtain the same spectral shape above 1 
keV as that given by {\tt COMPPS} model. The optical depth derived from 
simulations is usually smaller than the one derived from the {\tt COMPPS} model 
by about 30--40 per cent.

\begin{figure}
\includegraphics[width=\columnwidth]{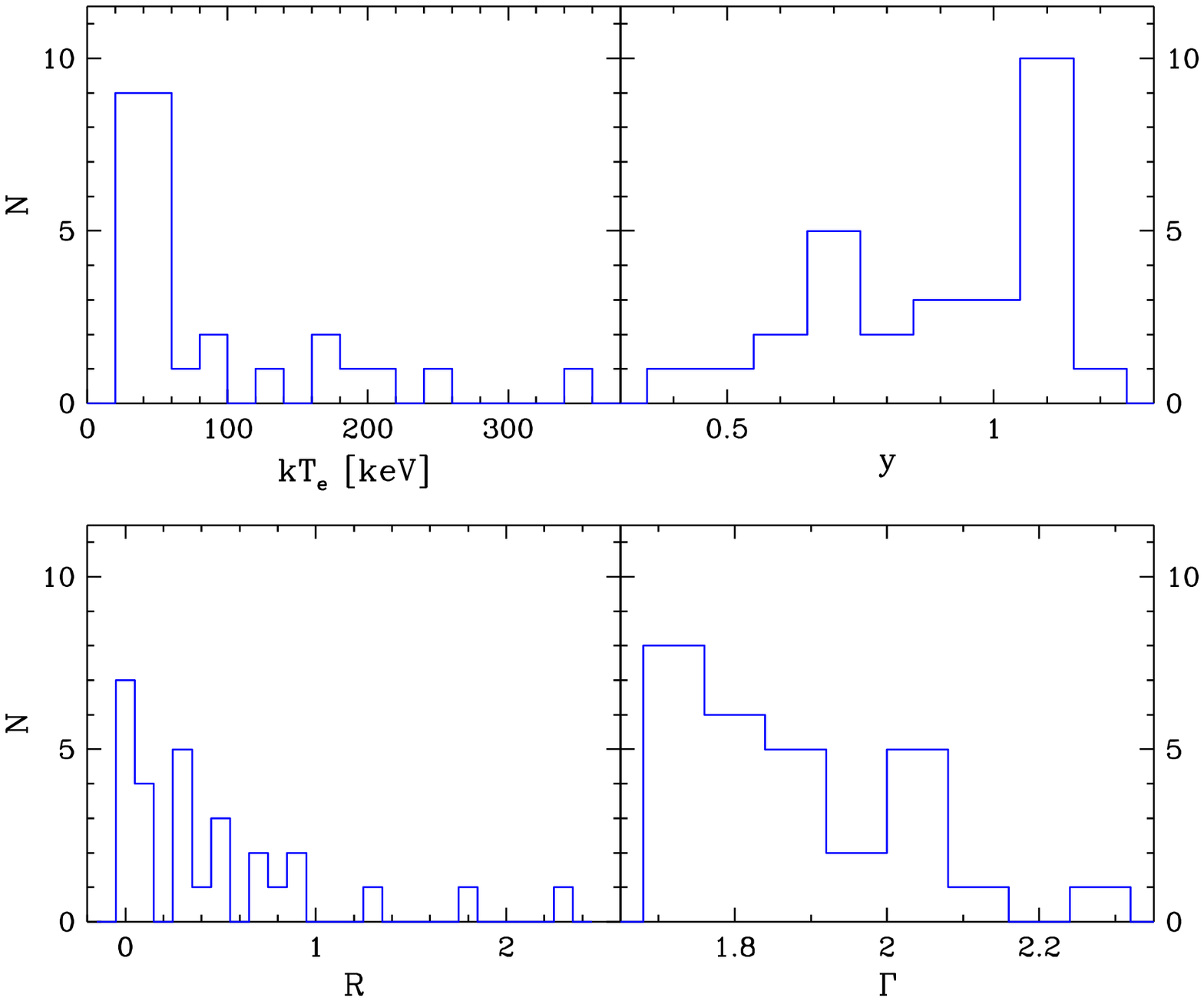}
\caption{Distributions of the best model values of the plasma temperature 
$kT_{\rm e}$, Compton parameter $y$, Compton reflection $R$ and spectral index 
$\Gamma$.}
\label{resdis}
\end{figure}

\begin{table}
\setlength{\tabcolsep}{1.3mm}
\centering
\caption{Median values of four main parameters of the continuum model for the 
studied sample.  The medians were computed for different model options and for 
the best selected model for the whole sample and for three Seyfert types. 
The uncertainties were computed as the lower and upper quartiles.} 
\label{averages}
%\footnotesize{
\begin{tabular}{lcccc}
\hline
Model option & $kT_{\rm e}$ [keV] & $y$ & $\Gamma$ & $R$ \\
\hline
A  & 64$_{-24}^{+136}$ & 0.83$_{-0.16}^{+0.23}$ & 1.89$_{-0.10}^{+0.10}$ &
0.34$_{-0.18}^{+0.53}$ \\[2pt] 
Ba & 52$_{-16}^{+265}$ & 1.00$_{-0.20}^{+0.15}$ & 1.77$_{-0.05}^{+0.09}$ &
0.81$_{-0.28}^{+0.77}$ \\[2pt] 
B & 50$_{-16}^{+135}$ & 0.92$_{-0.21}^{+0.15}$ & 1.83$_{-0.06}^{+0.15}$ &
0.29$_{-0.18}^{+0.35}$ \\[2pt] 
C & 49$_{-12}^{+105}$ & 0.89$_{-0.14}^{+0.20}$ & 1.88$_{-0.12}^{+0.12}$ &
0.25$_{-0.23}^{+0.37}$ \\[2pt] 
D & 51$_{-16}^{+96}$ & 0.90$_{-0.19}^{+0.18}$ & 1.82$_{-0.06}^{+0.14}$ &
0.30$_{-0.28}^{+0.30}$ \\[2pt] 
\hline
Whole sample 
& 48$_{-14}^{+57}$ & 0.94$_{-0.22}^{+0.16}$ & 1.81$_{-0.05}^{+0.18}$ &
0.32$_{-0.28}^{+0.33}$ \\[2pt] 
Type 1 
& 45$_{-11}^{+80}$ & 0.94$_{-0.23}^{+0.15}$ & 1.85$_{-0.08}^{+0.17}$ &
0.33$_{-0.15}^{+0.24}$ \\[2pt] 
Type 1.5 
& 44$_{-8}^{+70}$ & 1.04$_{-0.32}^{+0.04}$ & 1.80$_{-0.03}^{+0.21}$ &
0.30$_{-0.17}^{+0.17}$ \\[2pt] 
Type 2 
& 45$_{-29}^{+83}$ & 0.93$_{-0.21}^{+0.18}$ & 1.80$_{-0.05}^{+0.15}$ &
0.32$_{-0.25}^{+0.46}$ \\ 
\hline
\end{tabular}                                                                   
%}                                                                               
\end{table}                                                                    

Figure \ref{resdis} presents the distributions of values of the three main 
parameters of the continuum model and $\Gamma$ values found for the whole
sample with selected best-fitting model. The distributions of $kT_{\rm e}$, $R$ and 
$\Gamma$ values are asymmetric and do not resemble normal distributions. 
The values of the Compton parameter form a relatively broad peak centered around
0.8, overimposed by a narrow peak around $y$ = 1.1. Using the results of different 
spectral model options we have checked how the main parameters change with the 
model complexity. Since the distributions for all fitted models are asymmetric, 
to compare them we computed the median values instead of the standard averages. 
These numbers are shown in Table \ref{averages}. Although the sample is small
we can draw some conclusions. In 
general, more realistic models (B,C,D) lead to a smaller scatter of the 
$kT_{\rm e}$ and $R$ values, reducing the high-value tail of their distributions. 
Except for $R$ in the case of model Ba, the median values for all parameters are 
comparable. As shown in the lower part of Table \ref{averages}, the median 
values for all four parameters are quite similar for Seyferts type 1, 1.5 and 2.

\begin{figure}
\includegraphics[width=\columnwidth]{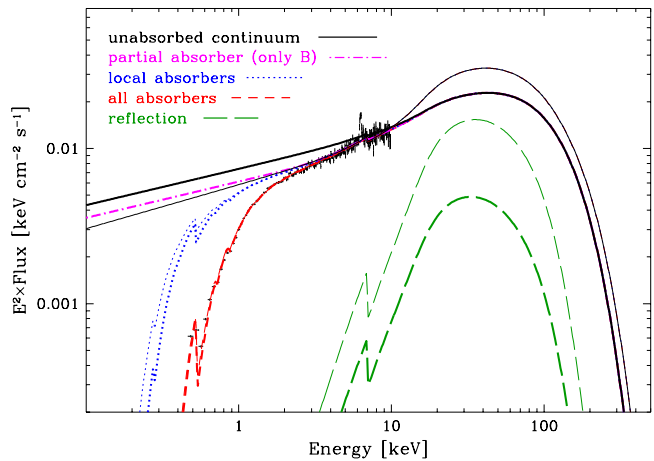}
\caption{Spectral components fitted to the \xmme spectrum of 4U~0517+17 with 
the models B (thick lines) and Ba (thin lines). Modifications of the primary 
continuum by all, local to the host galaxy and partially covering absorbers 
(only model B) are shown. The Compton reflection component is strengtened in 
model Ba.}
\label{bamodel}
\end{figure}

To understand the systematic differences between the results of models Ba 
and B we have studied the plots showing each spectral component 
separately. Figure \ref{bamodel} illustrates such an analysis for 
4U~0517+17. The bias introduced by model Ba can be explained as follows. 
If in the host galaxy there exists an ionized, partially covering 
absorber besides the fully covering absorber, a simple absorption model alone 
cannot reproduce a slight modification of the absorbed continuum due to the 
partial absorption. This modification present in model B is shown in Fig. 
\ref{bamodel} with the magenta curve. The missing partial absorber component in 
model Ba is counterbalanced by the fit of a strengtened reflection component. 
At the same time the fitted plasma temperature and Compton parameter can be 
altered due to the adjustment of several spectral components to match the strong
reflection component.

\subsection{Bolometric luminosity}
\label{bolum}

The bolometric luminosity is one of the most fundamental quantities characterizing 
radiation from a celestial object. In the case of AGN this is a particularly 
important parameter allowing to estimate the efficiency of the accretion 
process. An estimate of the bolometric luminosity is difficult 
when the spectral energy distribution is not sufficiently covered by measurements
(e.g., UV and very soft X-ray emission) and contamination arising from the host 
galaxy can influence the measurement of the AGN core flux. Thus, usually one uses some 
substitute based on the data from a given band. The X-ray band is preferred for 
AGN, as there is a negligible contamination from the rest of the host galaxy and 
the X-ray flux correlates well with the flux emitted by the AGN in the other 
energy bands (e.g., radio, infrared and optical). Commonly applied are luminosity 
estimates based either on the medium energy X-rays, e.g., that have been proposed 
by \citet{Marconi2004} for the 2--10~keV band, or on the hard X-rays, e.g., 
20--100~keV band chosen by \citet{Beckmann2009}. On the other hand, there are 
attempts to determine the bolometric luminosity from the entire spectral energy 
distribution (SED) seen for a given source (e.g., \citealt{Vasudevan2010}). 
Unfortunately, even very careful analysis of that type can suffer from some 
missing information when a large number of objects is considered. Very often 
this missing information pertains to the soft $\gamma$-ray emission above 100 
keV, where some AGN radiate a significant fraction. 

For the sake of taking into account the entire emission in the X-ray and 
$\gamma$-ray bands, we have chosen yet another estimate of the bolometric 
luminosity $L_{\rm bol}$ for our sample. Namely, we assumed that the luminosity 
integrated in the 1~eV -- 1~MeV band for the best-fitting unabsorbed model 
corresponds to half of the bolometric luminosity. Our $L_{\rm bol}$ values are 
listed in Table \ref{lumino} together with the luminosities computed for the 
seed photons and Comptonized continuum components. The last column of Table 
\ref{lumino} presents our estimates of the average Eddington ratio 
$\lambda_{\rm E}$ for each object. To compute $\lambda_{\rm E}$ we used the 
relation $L_{\rm Edd}$ = 1.3$\times 10^{38} M_{\rm BH}$ ergs s$^{-1}$ for the 
Eddington accretion luminosity, where the BH mass is given in solar mass units.

\begin{figure*}
\includegraphics[width=120mm]{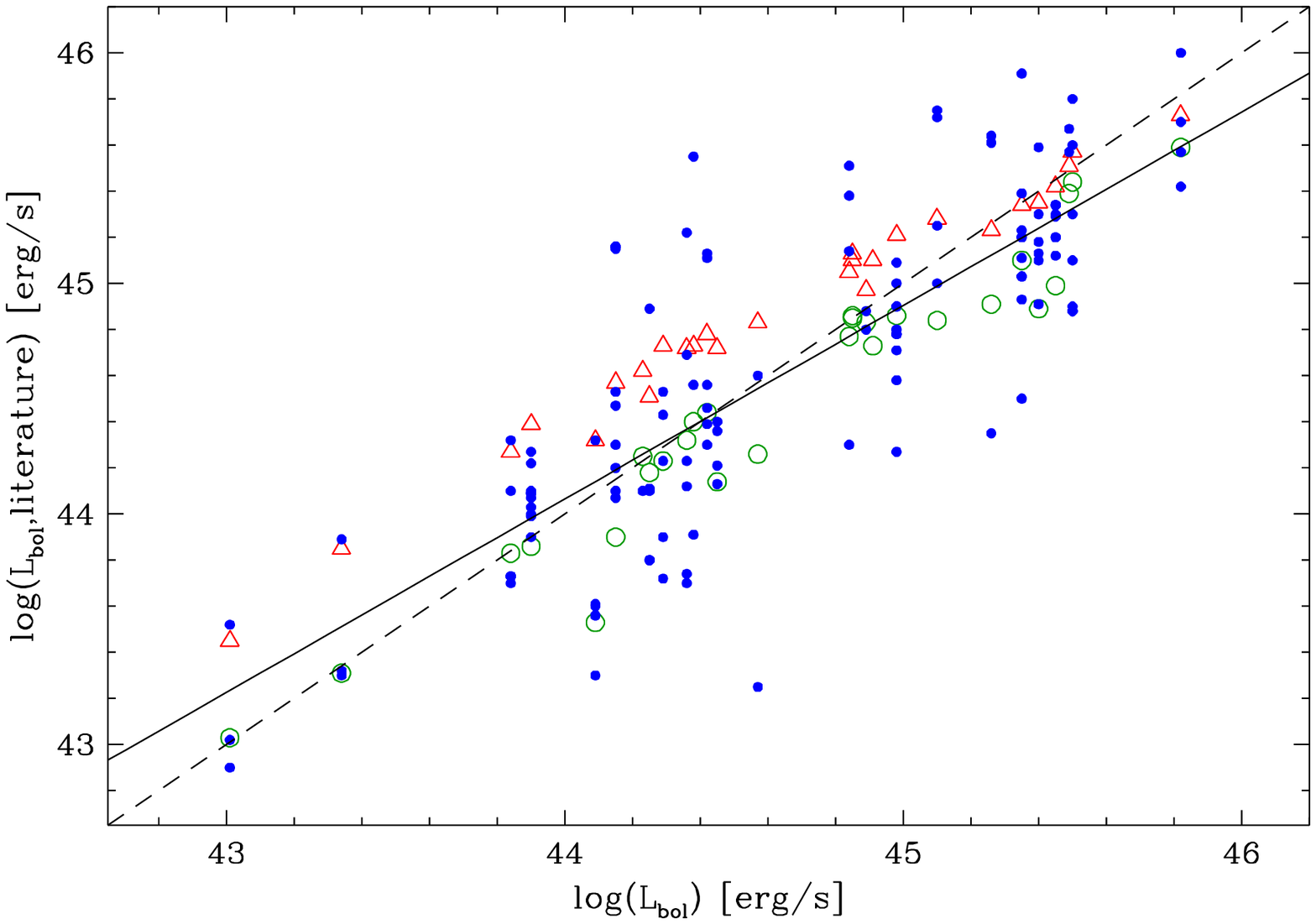}
\caption{Bolometric luminosity estimates for the studied AGN sample. Values 
taken from the literature are plotted against our bolometric luminosity measure,
$L_{\rm bol} = 2\times L$(1~eV -- 1~MeV). The data, shown with dots, come from: 
\citet{Khorunzhev2012}, \citet{Winter2012}, \citet{Vasudevan2009,Vasudevan2010}, 
\citet{Woo2002}, \citet{Wang2007}, \citet{Middleton2008}, \citet{Chitnis2009}, 
\citet{Zhang2011}, \citet{Zhou2010}, \citet{Raimundo2012}, 
\citet{Marinucci2012}, \citet{Bian2007} and \citet{Alonso2011}. Two other 
estimates are: $6\times L$(20~keV -- 100~keV), \citet{Beckmann2009} (circles), 
\citet{Marconi2004} formula using the 2--10~keV luminosity (triangles). Solid 
line shows the linear fit to the literature data. Dashed line is the equality 
line.}
\label{bolumi}
\end{figure*}

In order to investigate the accuracy of our estimates, we collected various 
luminosity estimates from the literature, based on different measurements and 
methods. In addition, we also computed the bolometric luminosity using the 
2--10~keV and 20--100~keV band fluxes integrated for the best-fitting model and 
the formulae of \citet{Marconi2004} and \citet{Beckmann2009}, respectively. All 
those luminosity estimates are plotted in Figure \ref{bolumi} as a function of 
our $L_{\rm bol}$ estimates. The dashed line in this figure indicates the 
one-to-one relation between the two luminosity estimates. Clearly, our estimates 
do correspond well to the average luminosity estimates for a given object, as 
determined with the other methods. Only for the low and high luminosity ends of 
the sample we observe a small departure from this trend. To quantify those 
deviations we fitted a line to the literature data in a function of our 
$L_{\rm bol}$, assuming that all literature data are affected by the same error.
The resulting relation is
\begin{equation}
\label{EQ2}
\log{L_{\rm bol,literature}} = (0.84\pm0.05)\times \log{L_{\rm bol}} + (7.1\pm2.3),
\end{equation}
the corresponding curve is plotted in Fig. \ref{bolumi} with a solid line.

An assumption of equal uncertainty for all literature data is the simplest
solution in situation when majority of those data lack the uncertainty estimate.
We have found that to obtain reduced $\chi^{2}$ = 1 for the above fit, a 0.4 dex 
error has to be set for each $\log{L_{\rm bol,literature}}$ value. For 
one-to-one relation the reduced $\chi^{2}$ = 1.07. The differences between the
two lines shown in Fig. \ref{bolumi} are much smaller than the scatter of the 
luminosity estimates seen in Fig \ref{bolumi}. Therefore, for the studied sample 
we keep our simple choice for $L_{\rm bol}$ estimate. We would like to stress 
that our choice was proved to be valid only for this particular AGN sample. 
A general relation similar to Eq. \ref{EQ2} should be verified with a larger AGN 
sample and with a reliable estimate of the literature data uncertainties.

Our $L_{\rm bol}$ luminosity estimate is twice the luminosity of the high 
temperature plasma. Interestingly, in a recent study of the luminosity of a 
sample of local Seyfert galaxies it was found that the corona and accretion 
disc intrinsically radiate approximately equal power \citep{Sazonov2012}. 
There are several other sources of the AGN radiation: jet, torus and outflows 
(e.g., those forming a broad emission line region, BLR). However, when a 
conversion of the gravitational energy into radiation is considered for a given 
accretion system, radiation reprocessed in torus should not be taken into 
account in the energy budget. The bolometric luminosity estimates of 
\citet{Vasudevan2010} and \citet{Sazonov2012} follow these lines, considering 
only a primary disc and plasma radiation. Since the disc emission in the UV and 
soft X-ray bands is hardly observed directly due to absorption, their estimate 
of the disc luminosity is indirect, based on the infrared emission from the 
torus illuminated by the disc. We used a simpler way to estimate the disc 
luminosity but our results are very similar to those of \citet{Khorunzhev2012}, 
used by \citet{Sazonov2012}. For 17 common objects we obtain a line with a slope 
0.95 and total $\chi^2$ = 6.6, assuming 0.4 dex uncertainty.

Whereas the emission in the infrared band is the torus reprocessed emission 
from the disc and plasma region, a radio emission induced by the jet kinetic 
energy can be important for the direct accretion power budget. Radio emission 
from the radio-quiet Seyferts is negligible compared to the disc and plasma 
emission. Among radio-loud objects in our sample only Cygnus A emits a large 
fraction of its energy in the radio band, approximately about 20\%, according to 
the spectral energy distribution (SED) data collected in the NASA Extragalactic 
database (NED). For the rest of the radio-loud objects in our sample similar 
estimate gives at most 1\% for 3C 390.3 and 3C 120, whereas for 3C 111 it is 
well below 1\%. Thus, only for Cygnus A there can be missing some fraction of 
the jet luminosity but our $L_{\rm bol}$ estimate (the most luminous object in 
Fig. \ref{bolumi}) do not show too low luminosity compared to the other results.
 
\begin{table}
\setlength{\tabcolsep}{1.3mm}
\centering
\caption{Logarithms of bolometric luminosities (in ergs s$^{-1}$) and the 
Eddington ratio $\lambda_{\rm E}$ for studied Seyferts computed for the selected 
best-fitting model. $L_{\rm seed}$ - luminosity of the seed photons component, 
$L_{\rm Comp}$ - luminosity of the Comptonized component of the spectrum 
integrated in the 1~eV -- 1~MeV band, $L_{\rm bol}$ - bolometric luminosity of 
the AGN, computed as a doubled value of the total model luminosity in the 1~eV 
-- 1~MeV band.}
\label{lumino}
\footnotesize{
\begin{tabular}{lcccl}
\hline
Object & $L_{\rm seed}$ & $L_{\rm Comp}$ & $L_{\rm bol}$ & $\lambda_{\rm E}$ \\
\hline
IC 4329A        & 43.65 & 44.67 & 44.98 & 0.070 \\
IGR J21247+5058 & 43.24 & 44.52 & 44.85 & 0.086 \\   
GRS 1734-292    & 43.77 & 44.55 & 44.91 & 0.0072 \\
NGC 4593        & 42.37 & 43.60 & 43.90 & 0.045 \\
4U 0517+17      & 42.81 & 44.03 & 44.38 & 0.122 \\
Akn 120         & 44.67 & 45.08 & 45.40 & 0.128 \\
ESO 141-55      & 44.17 & 44.79 & 45.26 & 0.352 \\    
3C 111          & 44.15 & 45.16 & 45.49 & 0.217 \\    
\hline
NGC 4151        & 42.24 & 43.51 & 43.84 & 0.020 \\
MCG+8-11-11     & 43.94 & 44.80 & 45.10 & 0.092 \\
Mrk 509         & 43.65 & 44.78 & 45.35 & 0.114 \\
4U 1344-60      & 42.70 & 43.87 & 44.23 & 0.028 \\
3C 120          & 44.64 & 45.13 & 45.45 & 0.352 \\
NGC 6814        & 41.73 & 43.04 & 43.34 & 0.012 \\
3C 390.3        & 44.04 & 45.20 & 45.50 & 0.072 \\
MCG-6-30-15     & 43.28 & 43.76 & 44.09 & 0.131 \\
\hline
NGC 4388        & 42.79 & 44.05 & 44.36 & 0.157 \\
NGC 2110        & 42.70 & 43.89 & 44.25 & 0.0084 \\ 
NGC 4507        & 42.75 & 44.12 & 44.42 & 0.039 \\
MCG-5-23-16     & 43.45 & 44.11 & 44.45 & 0.077 \\ 
NGC 5506        & 42.98 & 43.83 & 44.15 & 0.053 \\ 
Cygnus A        & 44.74 & 45.49 & 45.82 & 0.020 \\ 
NGC 5252        & 43.27 & 44.54 & 44.84 & 0.018 \\
ESO 103-35      & 43.55 & 44.23 & 44.57 & 0.176 \\
NGC 788         & 42.80 & 43.98 & 44.29 & 0.053 \\
NGC 6300        & 41.59 & 42.55 & 43.01 & 0.018 \\
NGC 1142        & 43.51 & 44.48 & 44.89 & 0.024 \\
LEDA 170194     & 43.26 & 44.55 & 44.85 & 0.014 \\ 
\hline
\end{tabular}                                                                   
}                                                                               
\end{table}    

\subsection{Compton amplification and Compton reflection}

\begin{figure}
\includegraphics[width=\columnwidth]{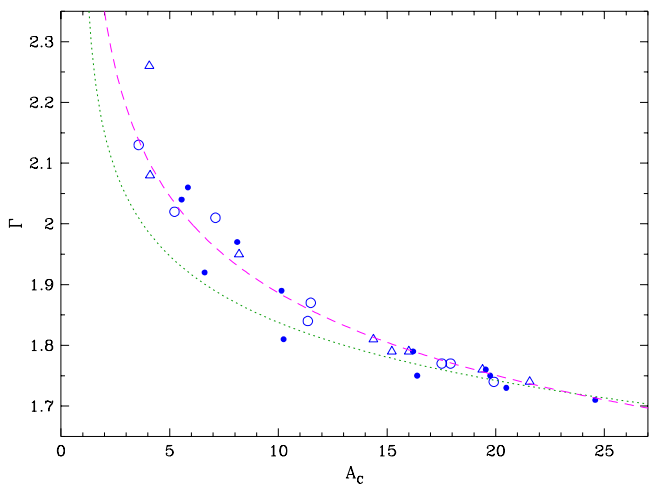}
\caption{Photon index $\Gamma$ as a function of the Compton amplification 
parameter $A_{\rm C}$ (circles - Sy~1, triangles - Sy~1.5, dots - Sy~2). Dotted and 
dashed lines show the power-law approximation of the $\Gamma$($A_{\rm C}$) 
dependence found by \citet{Beloborodov1999a} and \citet{Malzac2001}, 
respectively.}
\label{campli}
\end{figure}

There are several other parameters characterizing the AGN accreting system that
can be derived from the parameters of the models fitted to the X/$\gamma$ 
spectra. One of them is the Compton amplification factor $A_{\rm C}$, computed as 
the ratio of the Comptonized plus seed photon fluxes to the seed photon flux. 
We have computed $A_{\rm C}$ values using the model fluxes of the Comptonized and 
black-body components, integrated in the 1~eV -- 1~MeV band. In Figure 
\ref{campli} we present the spectral index $\Gamma$ plotted against $A_{\rm C}$.
This dependence can be well approximated by a power-law formula. As can be seen, 
our results agree particularly well with the approximation found for AGN by 
\citet{Malzac2001}, based on simulations of a dynamic corona atop an accretion
disc. 

\begin{figure}
\includegraphics[width=\columnwidth]{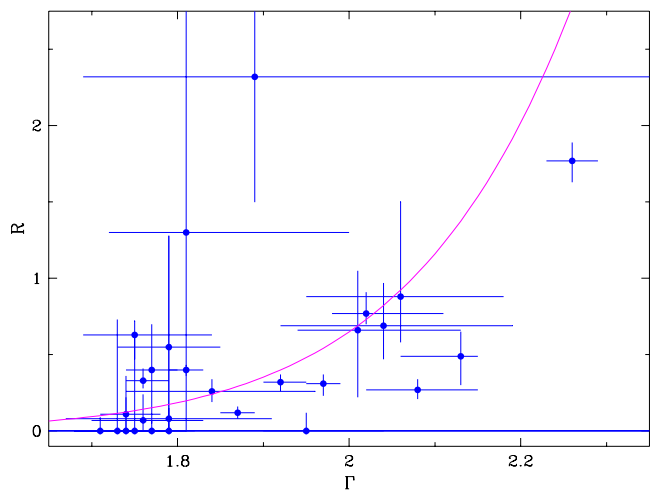}
\caption{Reflection strength $R$ as a function of the photon index $\Gamma$
determined for our sample. Line shows the power-law approximation of the 
$R$($\Gamma$) dependence found by \citet{Zdziarski1999}.}
\label{gammar}
\end{figure}

A detailed study of the correlations between our results and other parameters 
characterizing accretion systems at the centre of Seyfert galaxies is out of 
the scope of the present paper. The only dependence we want to refer to is the     
correlation between the Compton reflection and photon index found for Seyferts 
and BH binaries in many previous studies (e.g., those based on the Ginga data, 
\citealt{Zdziarski1999}). Figure \ref{gammar} presents our $\Gamma$-$R$ data 
compared with the relation found in \citet{Zdziarski1999} (this relation takes 
into account an internal correlation of these two parameters in the {\tt PEXRAV}
model used for Ginga spectra). Although the broad-band spectral analysis 
presented here allows for a better determination of both $\Gamma$ and $R$, the 
true correlation can be suppressed due to the averaging of data collected during 
various spectral states. This explains a less prominent $R$-$\Gamma$ correlation 
than that seen for the short exposure time observations reported in 
\citet{Zdziarski1999}. 

\section{Discussion}
\label{sedis}

\subsection{Comparison with previously published results}
\label{codis}

\subsubsection{Comptonization model, \groe and \saxe samples}

First we compare our results with those quoted in Sec. \ref{intro}, namely
diverging $kT_{\rm e}$ estimates based on \groe \citep{Zdziarski2000} and \saxe 
\citep{Petrucci2001} observations. Analysis of the \groe data is compatible with 
ours thanks to the application of the same Comptonization model. The \gro/OSSE 
spectra of Type~1 (17 objects) and Type~2 (10 objects) nuclei were stacked 
separately and fitted with a fixed $R$ = 0.75. Best-fit Compton parameter and 
plasma temperature are: 0.89$_{-0.52}^{+0.36}$ and 69$_{-28}^{+134}$~keV for Sy~1, 
and 1.09$_{-0.41}^{+0.29}$ and 84$_{-31}^{+101}$~keV for Sy~2. Uncertainty of 
those estimates is quite large, despite the high quality of the OSSE spectra, 
due to the lower limit of the energy band at 50~keV. The agreement with our results, 
presented in Table \ref{averages} and Fig. \ref{resdis}, is very good. Thus our 
analysis of the Seyfert spectra is fully consistent with the results obtained 
for the OSSE detector.

Analysis done with the \saxe spectra \citep{Petrucci2001} was limited to only
six sources, among them IC~4329A, NGC~4151, ESO~141-55 and Mrk~509 from our 
sample. The authors used the anisotropic Comptonization model of 
\citet{Haardt1993} assuming a slab geometry for the corona above the disc. The 
slab geometry gives typically $kT_{\rm e}$ values similar to those derived for the 
spherical plasma and $y$ values about two times smaller (see e.g., 
\citet{Lubinski2010}). The 
$kT_{\rm e}$ values quoted in \citet{Petrucci2001} are in the range between 170 and 
315~keV, with quite small uncertainties (typically 5--30~keV). Such a high 
accuracy is rather surprising because the sensitivity of the \sax/PDS detector 
was only slightly higher than that of ISGRI and the \saxe exposure times were 
about ten times shorter than the ISGRI exposure times. The Compton reflection 
estimated with the \saxe spectra is also quite large, typically close to $R$ = 1 
with an error of about 10--30 per cent. All the objects with high $kT_{\rm e}$ estimates 
based on the \saxe data belong to the low $kT_{\rm e}$ group of our analysis, with 
the $R$ $\ll$ 1. Therefore, there is a substantial discrepancy between these two 
analyses. Higher $kT_{\rm e}$ estimates based on the \saxe spectra can be explained 
by the fact that the best-fitting model are driven by the low-energy spectra from 
the LECS and MECS detectors, having many channels with relatively small errors, 
whereas the PDS spectra are quite scarce above 50 keV. Our suggestion is based 
on inspection of figure~2 in \citet{Petrucci2001}, where the high-energy tails 
of the IC~4329A and especially NGC~4151 spectra are softer than the plotted 
model. Taking into account the quality of the \saxe spectra above 100~keV it 
seems that the computed $kT_{\rm e}$ uncertainties are underestimated in situation
where the best-fitting $kT_{\rm e}$ is much higher than the upper energy limit of the 
fitted spectrum. In turn, the \saxe $R$ values are increased due to a higher 
plasma temperature making the reflected component flatter.

\subsubsection{Phenomenological model, large samples}

A detailed comparison with all previously published results of the spectral 
modelling done already with various X/$\gamma$ spectra of the studied sample 
is out of scope of the present paper. We limit this analysis to several 
recent papers presenting the results obtained for large samples of Seyfert 
galaxies and several papers on individual objects. Spectra of all Seyfert 
samples discussed here were analysed with a phenomenological model in the form 
of a power-law with a high energy cut-off $E_{\rm C}$ (XSPEC model {\tt PEXRAV}). 
Hard X-ray spectra fitted by those groups were taken with \sax/PDS + low-energy 
spectra from \sax/MECS 
\citep{Dadina2007}, \xte/HEXTE + \xte/PCA \citep{Rivers2013}, \swift/BAT +
\suzakue and \xmme \citep{Winter2012} and \integral/ISGRI: + \integral/JEMX
\citep{Beckmann2009}, + \xmm, \chandra, \ascae and \swift/XRT 
\citep{Molina2009}, + \xmme \citep{DeRosa2012} and + \xmme and \swift/BAT
\citep{Malizia2014}. All those analyses applied a fully covering neutral 
absorber, whereas a partial/ionized absorption component was added only by 
\citet{Dadina2007}, \citet{Molina2009} and \citet{Malizia2014}.   

\begin{figure*}
\includegraphics[width=\columnwidth]{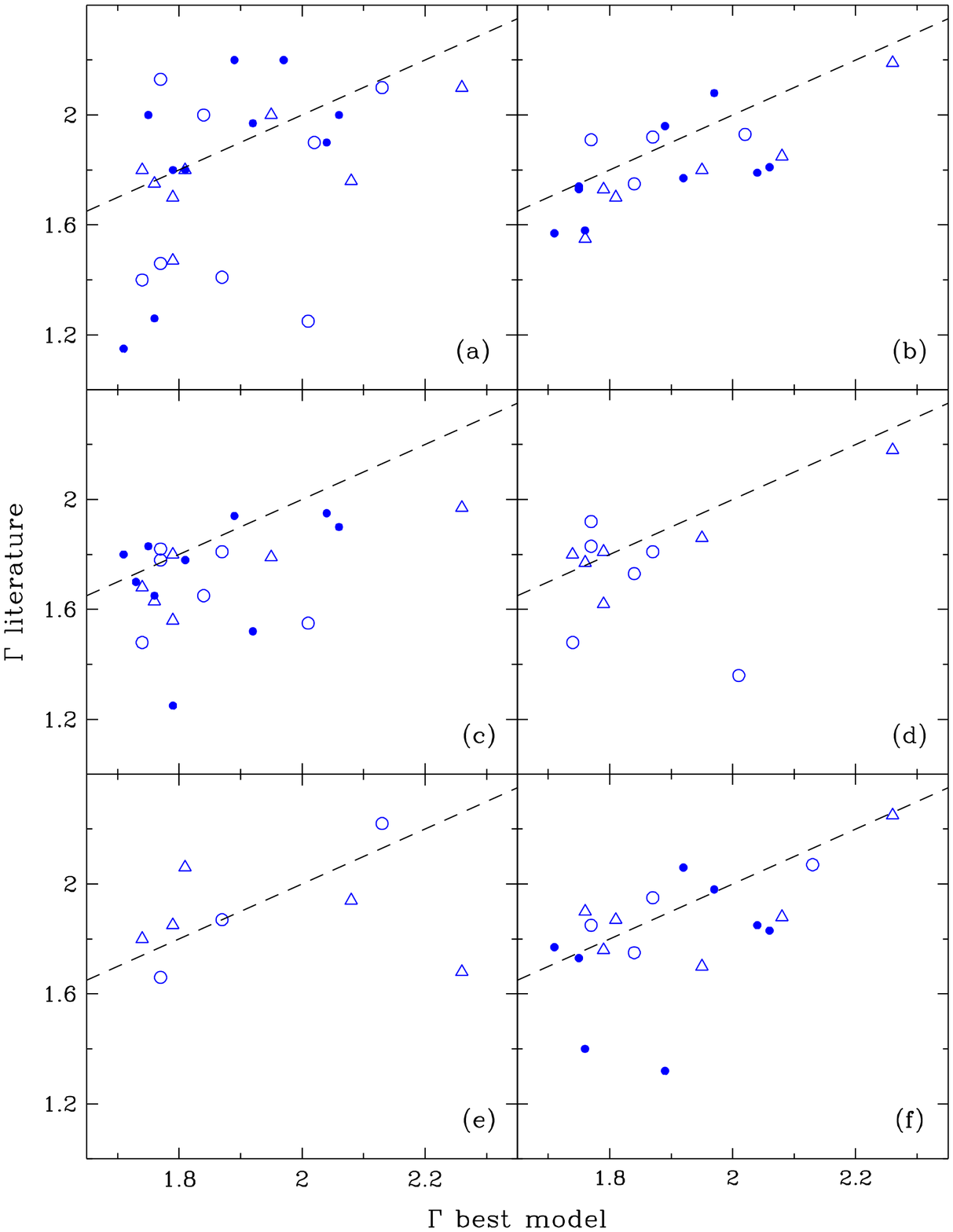} \hskip16pt
\includegraphics[width=\columnwidth]{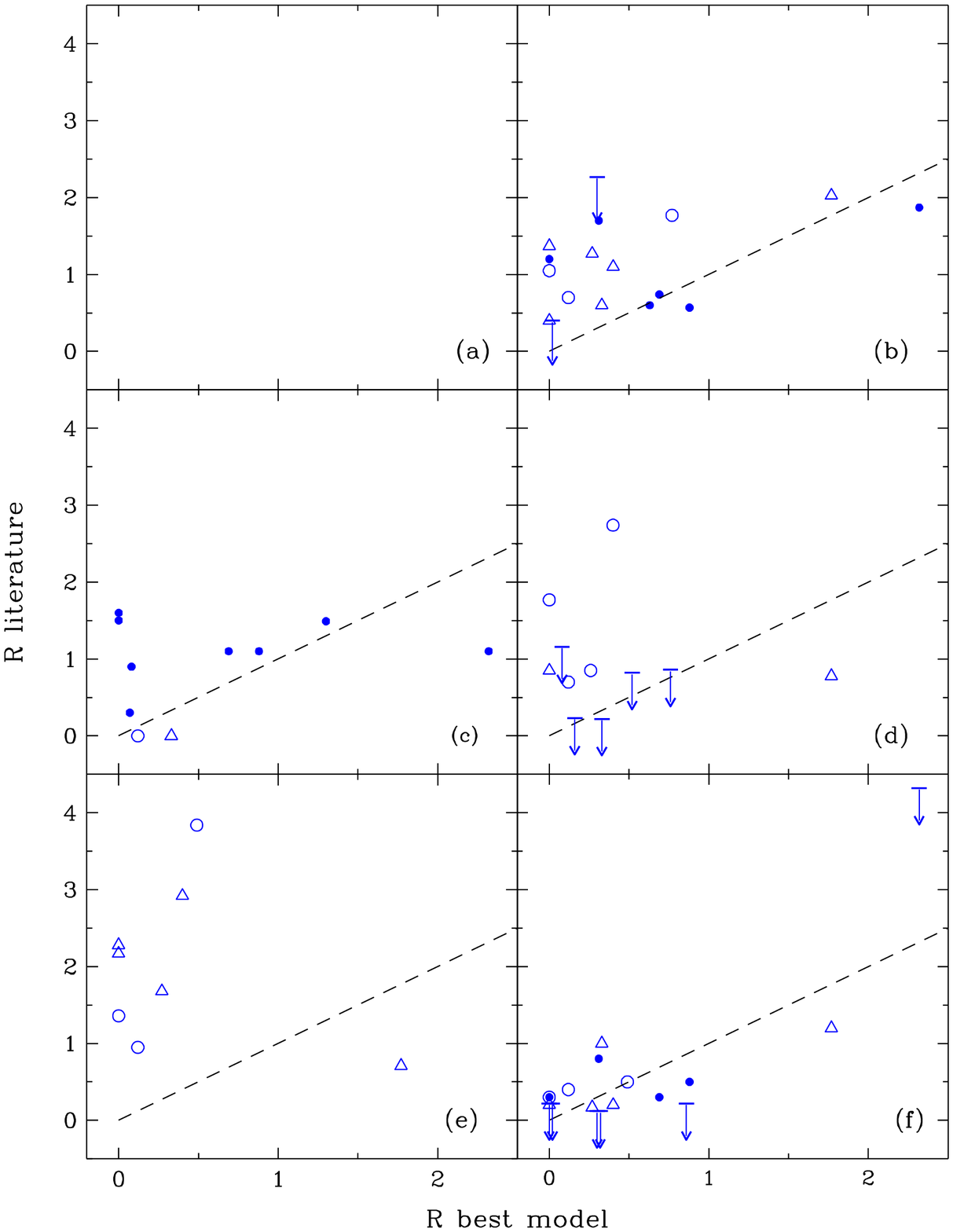}
\caption{Left: Comparison of the photon index $\Gamma$ determined in several 
recent analyses with our $\Gamma$ results. Right: Comparison of the Compton 
reflection strength $R$ determined in several recent analyses with our $R$ 
results (there is no $R$ data in \citet{Beckmann2009}, panel a). Data were taken 
from: (a) - \citet{Beckmann2009}, (b) - \citet{Dadina2007}, (c) - 
\citet{Malizia2014,DeRosa2012}, (d) - \citet{Molina2009}, (e) - 
\citet{Winter2012}, (f) - \citet{Rivers2013}. Object types: circles - Sy~1, 
triangles - Sy~1.5, dots - Sy~2. The dashed lines show the equality lines.}
\label{otgam}
\end{figure*}

In Figure \ref{otgam} we compare the results of the studies quoted above with 
our results for $\Gamma$ and $R$, respectively. The largest AGN sample 
considered here is the \integrale catalogue of \citet{Beckmann2009}; their 
analysis is simplified and suffers also from a lack of high-quality low-energy 
spectra (\integral/JEMX has rather limited sensitivity). Therefore, the obtained 
$\Gamma$ values are quite scattered (see Fig. \ref{otgam}, panel (a)), and the 
authors did not fit a Compton reflection component. An analysis of the \saxe 
spectra by \citet{Dadina2007} provided $\Gamma$ results consistent with ours 
within $\Delta\Gamma = \pm 0.1$ but the $R$ values are usually larger than ours 
(panels (b)). For the rest of the $\Gamma$ comparisons (panels c--f) there is a 
relatively large scatter of values observed. On average, in all cases (except 
for \citet{Winter2012}) our spectra seem to be systematically softer, in 
agreement with the results of the tests presented in Sec. \ref{selmod}: harder 
spectra are fitted with simpler absorption model Ba. 

Interestingly, our results show the narrowest range of the fitted $\Gamma$ 
values, with a sharp cut-off at low values. The shape of the $\Gamma$ 
distribution is the result of the $y$ and $kT_{\rm e}$ distributions, both with 
a relatively narrow main peak. When the spectra are fitted with the COMPPS 
model, the spectral slope is a result of Comptonization of seed photons. Since 
the seed photon distribution must occupy some region within optical, UV and soft 
X-ray bands, the Comptonized spectra cannot be too hard because otherwise the 
normalization of the seed photon distribution will become too small to produce a 
sufficient number of Compton upscattered photons for a given optical depth of 
the plasma. An additional regulation of the spectral slope comes from the fitted 
kTe, due to the dependence of the Comptonization efficiency on the energy gain 
per one scattering. Therefore, Comptonization model provides physical 
limitations on the spectral slope, when compared to the phenomenological models 
where the $\Gamma$ parameter is completely free. 

Comparison of the fitted $R$ values shown in Fig. \ref{otgam} confirms our 
conclusion that simpler models produce, on average, a larger Compton reflection.
Only the results of the \xtee data analysis by \citet{Rivers2013} are roughly 
consistent with our results. We have also compared the high-energy cut-off 
values $E_{\rm C}$ fitted with the {\tt PEXRAV} model with our $kT_{\rm e}$ 
results, using two relations $kT_{\rm e} = E_{\rm C}/2$ for $\tau \lesssim 1$ 
and $kT_{\rm e} = E_{\rm C}/3$ for $\tau \gg 1$ \citep{Petrucci2001}. A vast 
majority of the plasma temperature estimates based on the exponential cut-off 
model appeared to be larger than the $kT_{\rm e}$ results obtained by us with 
the Comptonization model. This difference can be explained again by less complex 
models applied in the analysis of larger samples. By less complexity of the 
model we mean several factors: a simpler absorption model, an exponential 
cut-off instead a Comptonized continuum (which decreases more rapidly at high 
energy, see Sec. \ref{secxb}) and a neglected variability of the low-energy 
spectra with respect to the spectral slope and absorption.

\citet{Ricci2011} presented an analysis of the average spectra extracted for 
several classes of AGN from the stacked \integral/ISGRI images. Interestingly,
they determined with the {\tt PEXRAV} model an average photon index of $\Gamma$ 
= 1.8, i.e., the same as our sample median. Their mean $E_{\rm C}$ $\lesssim$ 
200~keV is also compatible with the arithmetic mean $kT_{\rm e}$ $\approx$ 
80~keV for our sample, assuming $kT_{\rm e}-E_{\rm C}$ conversion factor between 
2 and 3. Moreover, for the three Seyfert groups (Sy~1, Sy~1.5, Sy~2) with 
$N_{\rm H}$ $<$ $10^{23}$ cm$^{-2}$ they found upper limits for $R$ in the range 
0.4--0.5, i.e., lower than usually reported and again compatible with our $R$ 
distribution median.   
  
\subsubsection{Short-term observations, individual objects}

Studies discussed so far in this Section were done for samples of AGN and the 
spectral properties were derived for hard X-ray spectra collected over several 
years, usually accompanied by the low-energy spectra with shorter exposure time. 
Only analyses of the \saxe \citep{Dadina2007} and \xtee \citep{Rivers2013} were 
done for contemporary low- and high-energy spectra. Thus it was desirable to 
compare our results with some results based on shorter but roughly contemporary 
broad-band observations done for individual objects. Although there can be some 
differences due to shorter sampling time, many of such observations were 
analysed with the Comptonization models, allowing us to directly compare the 
$kT_{\rm e}$ values. 

Since the sensitivity of the \nuste satellite in the 10--50~keV band is much 
higher than the sensitivity of any other satellite operating so far, we 
considered here mainly the results of the \nuste observations. Contemporary 
\suzakue and \nuste spectra of IC~4329A in a flux state near its long-time 
average were analyzed in detail by \citet{Brenneman2014}. Their plasma 
temperature and optical depth values, $50_{-3}^{+6}$ keV and $2.3\pm0.2$, 
respectively, found with the {\tt COMPTT} Comptonization model agree well with 
our results shown in Table \ref{mainres}. Similar overall agreement is found 
also for the high-temperature object Akn~120, which was observed simultaneously 
by \nuste and \xmme \citep{Matt2014}. The energy coverage of the \nuste 
detectors ($\lesssim$ 80~keV) precluded more precise determination of the 
high-energy cut-off value in this case. Both continuum models, {\tt PEXRAV} and 
{\tt COMPPS}, applied to fit the cut-off gave only a rough estimate, 
$E_{\rm C} > 340$~keV and $kT_{\rm e}$ in between 110 and 210~keV, respectively. 
Whereas $R$ = 0.26$\pm$0.08 found by \citet{Matt2014} with the {\tt PEXRAV} 
model is marginally consistent with our result (0.5$\pm$0.2), their $\Gamma$ 
= 1.79$\pm$0.03 is much smaller than our estimate (2.13$_{-0.07}^{+0.02}$). This 
discrepancy can be explained by a variability of the emission and a difference 
in the absorption model: there was no warm absorber found for the \nuste 
observation whereas our best fit needed a modest ionized, partly covering 
absorption. 

Another object from our sample with recently analysed \nuste spectra is NGC 
2110 \citep{Marinucci2015}. The \nuste spectra were fitted with a simple 
absorption model, accompanied by either a phenomenological or Comptonization 
model for the main continuum component. Comptonization model {\tt COMPTT} 
resulted in $kT_{\rm e}$ = 190$\pm$130~keV, i.e., close to our value of 230~keV. 
We note a large discrepancy for the Compton reflection strength, with the upper 
limit estimated at $R$ = 0.14 for both \nuste observations fitted with a cut-off 
power-law + reflection model, whereas our $R$ value is quite large, around 0.6.
Our fit with model Ba gave a similar $R$ value, thus the discrepancy is rather 
not related to a different absorption model. The {\tt PEXRAV} model accompanied 
by three partially covering absorbers was used to analyse two \suzakue 
observations of NGC 2110, with no reflection detected \citep{Rivers2014}. Thus, 
a possible explanation of the different $R$ values can be different continuum 
model used in our and other analyses. In addition, our \integrale NGC~2110 
spectra cover the period before the brightening of the source in 2012, whereas 
the two \nuste observations were performed in 2012 and 2013. There can be some 
change of the spectral properties of the source after unusual brightening. 
Indeed, when we added the recent \integrale data extracting a new summed ISGRI 
spectrum, the preliminary fit resulted in much smaller $R$ $\approx$ 0.25. 

Thanks to its excellent sensitivity the \nuste satellite will bring more 
interesting results for other objects from our sample, especially for those with 
a cut-off at low energy. Since \nuste observations are sampling the emission of 
a given source over a shorter time and less frequently, some difference in the 
results will always be expected. \integrale is more suited for the studies of 
the average emission over long periods, thus both these missions well complement 
each other. But even for \integrale we note clear differences for spectra taken 
over different periods. An example is Mrk~509, target of an extensive
multi-mission observation campaign in 2009 \citep{Kaastra2011}. A spectral 
analysis with two Comptonizing, hot and warm, coronae revealed their 
temperature around 100~keV and~1 keV, respectively \citep{Petrucci2013}. In our 
spectral analysis we modelled only the hot corona and found its $kT_{\rm e}$ = 
39$_{-9}^{+20}$~keV. In order to clarify this difference we have fitted our 
model to the ISGRI spectrum extracted only for the 2009 campaign period, 
accompanied by contemporary \xmme spectra. The best-fitting value for the plasma 
temperature $kT_{\rm e}$ = 82$_{-33}^{+42}$~keV is consistent with the value 
determined by \citet{Petrucci2013}, although it is less accurate than the result 
obtained using all available spectra. When all spectral data from \integrale are 
analyzed, the cut-off is clearly seen as shown in Fig. \ref{specs}. 

\subsection{Plasma diagnosis}
\label{pladi}

\subsubsection{Geometry and seed photons}

Contemporary models of the radiation processes in the accreting BH systems 
assume a hybrid, thermal plus non-thermal plasma 
\citep{Belmont2008,Veledina2011}. The non-thermal electrons are responsible for 
the synchrotron radiation cooling the plasma in addition to the emission from 
the cold accretion disc. Based on the low values determined for both 
$kT_{\rm e}$ and $R$ we propose a qualitative argument supporting the importance 
of the synchrotron cooling process. First, $R$ with the median value around 0.3 
indicates that the solid angle, covered by the accretion disc as seen from the 
plasma region, is relatively small, especially if one takes into account that 
some fraction of the reflection comes from the torus. Thus, if the disc and 
plasma are well separated spatially, we expect an even smaller solid angle under 
which the compact plasma region is seen from the disc. This means a reduced flux 
of the disc soft photons cooling the plasma. At the same time we observe low 
$kT_{\rm e}$ values for many low-$R$ objects. Therefore, there should exist 
another mechanism cooling the plasma and the synchrotron self-absorption is a 
plausible explanation. The range of Eddington ratios where the Comptonization of 
the synchrotron photons can be dominant, was suggested to be within 0.0001--0.01  
\citep{Malzac2012}. The majority of the low-$kT_{\rm e}$ and low-$R$ objects of 
our sample accrete close to that range, below $\lambda_{E}$ = 0.05. 

We have estimated the relative Compton parameter variability, $\Delta y/y$, 
where $\Delta y$ was computed as the standard deviation for the arithmetic mean. 
For this we used $y$ values fitted to the low-energy spectra with the best model 
selected by the Akaike test among those allowing for the spectral slope 
variation (option B or D). Except for NGC~4388, NGC~788 and NGC~1142 all object 
exhibit $\Delta y/y$ $\leq$ 20 per cent. Such a stability of the Compton 
parameter can be explained by the synchrotron boiler model, assuming a complete 
absorption of the synchrotron emission within the corona. Figure 3 of 
\citet{Ghisellini1998}, showing a dependence of $kT_{\rm e}$ and $y$ on the 
compactness of the injected non-thermal electrons, presents the $y$ values in a 
range very similar to that found for our sample, supporting the synchrotron 
self-Compton model. The plasma temperature shown in that figure is too high for 
a large range of compactness values but it can be shifted towards lower values 
when the other cooling radiation (disc photons) is taken into account. Recent 
studies of the synchrotron self-Compton model confirmed the stability of the 
spectral slope against variations of many parameters such as magnetization, 
optical depth or Eddington ratio \citep{Veledina2011}. Nevertheless, models 
including the Comptonization of the disc photons can also be adjusted to get a 
stable spectral slope \citep{Stern1995}, although in this case the geometry of 
the plasma and the disc should allow for their close interaction. Thus, taking 
into account that our $kT_{\rm e}$ and $R$ results point towards a separation of 
these two regions, the synchrotron regulation appears highly probable.

The fact that the $R$-$\Gamma$ correlation holds even for the mean values of 
these parameters indicates that both cooling processes are important. The 
self-regulating effect of the synchrotron self-Compton model can be smaller if 
some fraction of synchrotron emission escapes from the plasma. An extension of 
the \citet{Veledina2011} model was proposed to explain a complicated connection 
between the variability of the optical/infrared (OIR) and X-ray emission of the 
BH systems \citep{Veledina2013}. In this model most of the OIR emission 
originates in a hot flow, closer to the BH than the truncated accretion disc. 
The source of this emission is the synchrotron radiation of non-thermal electron 
population. Such a model, with the changing inner radius of the disc, offers 
essentially the same explanation for the $R$-$\Gamma$ correlation 
\citep{Poutanen2014} as that presented in \citet{Zdziarski1999}.
   
Non-thermal electrons are needed to get a substantial flux of synchrotron 
photons, ensuring an efficient plasma cooling down to the observed temperatures 
(especially if the magnetic field is relatively weak). As can be seen from the 
ISGRI spectra of Seyferts (e.g., Fig. \ref{sp4151}), we are still far from the
possibility of observing the high-energy power-law tails, such as those found 
in the spectra of Galactic BH binaries 
\citep{McConnell2000,Wardzinski2002,Zdziarski2012}. Therefore, a direct 
measurement of the non-thermal electrons fraction will not be possible for 
Seyferts with any current and, presumably, any near-future satellite. 

\subsubsection{Plasma radiation power}

Our estimate of the bolometric luminosity is based on an assumption that both 
the plasma and the disc emit approximately half of the total AGN radiation 
power. As mentioned in Sec. \ref{bolumi}, \citet{Sazonov2012} found a similar 
equality of the powers generated by the disc and by the hot corona. Their disc 
luminosity estimate is based on the mid-infrared luminosity of the torus, where 
the disc emission is reprocessed. For the hot corona they assumed that 30 per 
cent of its total flux is emitted in the 17--60~keV band. The 30 per cent
fraction corresponds to the cut-off power-law model with $\Gamma$ = 1.7 and 
$E_{\rm C}$ = 200~keV, consistent with the average spectrum of the local AGN 
sample studied by \citet{Sazonov2008}. The ratio of the observed disc luminosity 
to the corona luminosity found by \citet{Sazonov2012} is $\approx$ 2. Then, 
taking into account a reprocessing of the corona emission in the disc, they 
conclude that the intrinsic luminosities of the disc and the corona are roughly 
equal. Since $L_{\rm bol}$ computed by us is related to the observed, not 
intrinsic corona flux, there is a discrepancy between \citet{Sazonov2012} and 
our results. The lower corona luminosity found by \citet{Sazonov2012} can be 
explained by their assumption that their $L_{\rm HX}$ luminosity determined for 
the 17--60 keV band corresponds to $\approx$30 per cent of the total corona 
luminosity. As can be checked, this assumption is valid if their model is 
integrated in a limited range between 2 and 300 keV, whereas for the 0.001--1000 
keV band $L_{\rm HX}$ corresponds to only $\approx$21 per cent of the corona 
luminosity. Therefore, compared to our results their corona luminosity is 
underestimated by a factor $\approx$1.5. This explains why our $L_{\rm bol}$ 
estimate gives good results under an assumption of equal disc and corona 
luminosities.

\citet{Sazonov2012} do not propose an explanation for the equal powers of the 
disc and corona emission. Since they interpret their other results within a 
scenario of the compact corona seen at small solid angle from the disc, any 
disc-corona interaction mechanism leading to the emitted power equality can be
insufficient. For the alternative scenario, assuming an extended corona above 
the disc, there was such mechanism proposed for the Galactic black holes 
\citep{Schnittman2013}. Based on the magnetohydrodynamic simulations results 
they conclude `the MHD turbulence in an accretion disc can lead to dissipation 
outside the disc's photosphere strong enough to power hard X-ray emission 
comparable in luminosity to the disc's thermal luminosity'. However, this type 
of model cannot explain the small Compton reflection determined for Seyfert 
galaxies as well as positive lags between the hard X-ray and optical/UV 
emission. In addition, similarly to the other models assuming a corona extended 
above the disc, it predicts too steep X-ray spectra in the 2--10~keV band, with 
$\Gamma$ well above 2 (see fig. 12 in \citealt{Schnittman2013}).

\subsection{Comparison with Galactic BH binaries} 

The temperatures of the Comptonizing plasma found for Galactic black hole 
binaries observed in the hard state are typically in the range between 70 and 
100~keV, whereas during the soft state they are in the 10--40~keV range 
\citep{Malzac2012}. Both soft and hard state BHB and the low-temperature AGN 
samples are far too small to judge whether there is a significant difference 
between their mean $kT_{\rm e}$ values. The other parameter values, like 
$\Gamma < 2$ or $\lambda_{\rm Edd} < 0.05$, computed for the majority of our 
sample, correspond rather to the values typical for the hard state BHB. 
Moreover, as stated in \citet{Malzac2012}, the only model consistent with data 
for soft state BHB is the accretion disc extended down to the last stable orbit 
and small active coronal regions above the disc. Such a model will result in $R$ 
close to 1, not consistent with our results. This discrepancy can be solved by 
applying a dynamic plasma model, forming mildly relativistic outflows 
\citep{Beloborodov1999a} but this in turn demands a strong magnetic field needed 
for fast energy dissipation in the corona. Therefore, our results, consistent 
with the scenario of plasma and disc being well separated, support the hard 
state BHB - Seyfert correspondence. 

\subsection{Consequences for CXB studies}
\label{secxb}

\begin{figure}
\includegraphics[width=\columnwidth]{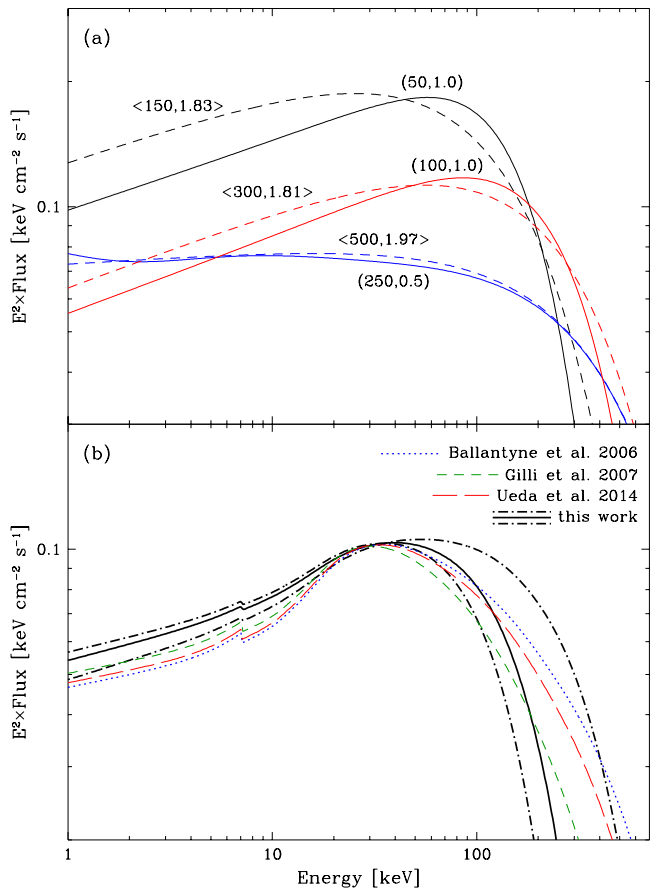}
\caption{(a) Comparison between Comptonization model ({\tt COMPPS}, solid lines) 
and exponentially cut-off power-law model ({\tt PEXRAV}, dashed lines) 
approximating it. $kT_{\rm e}$ and $y$ parameters of {\tt COMPPS} models are 
given in round brackets, $E_{\rm C}$ and $\Gamma$ parameters of {\tt PEXRAV}
models are given in angle brackets. Normalization is arbitrary, models do not 
include the Compton reflection component. (b) Examples of the {\tt PEXRAV} 
template spectra used in modelling of the AGN population synthesis compared with 
the Comptonization spectrum corresponding to the median values of $kT_{\rm e}$, 
$y$ and $R$ determined by us. The dash-dotted lines show models corresponding to 
the quartiles of the $kT_{\rm e}$ distribution determined by us. Spectra are 
normalized to the same level at 30~keV.}
\label{xrbspec}
\end{figure}

The $X/\gamma$ template spectra of AGN are one of the most important elements of
the population synthesis models aiming at explaining the CXB spectrum. Since 
there are many other constraints and effects to be taken into accout in such 
studies, for simplicity the template spectra are commonly approximated with an 
absorbed, exponentially cut-off power-law model, accompanied by a Compton
reflection (e.g., \citealt{Treister2009,Ueda2014}). However, an approximation of 
the high-energy cut-off of the Comptonized spectrum with the exponential cut-off 
is known to be poor, especially for the spectra with $kT_{\rm e} \lesssim$ 
100~keV (see fig. 5 in \citealt{Zdziarski2003}). To illustrate this effect again 
we have plotted in Fig. \ref{xrbspec} three Comptonization models together with 
the corresponding phenomenological models approximating them. Two Comptonization 
spectra were computed for $kT_{\rm e}$ = 50~keV, i.e., close to our sample 
median, the third spectrum is for $kT_{\rm e}$ = 250~keV, i.e., slightly above 
the highest more accurately determined temperature within our sample (NGC~2110). 
Cut-off energy $E_{\rm C}$ was adjusted according to the relation quoted in Sec. 
\ref{codis}. As can be seen in panel (a) of Fig. \ref{xrbspec}, for high 
$kT_{\rm e}$ the exponential cut-off with $E_{\rm C}$ = 2$kT_{\rm e}$ relatively 
well approximates the Comptonization cut-off. 

For the low $kT_{\rm e}$ regime the situation is much worse: for both 
$kT_{\rm e}$ = 50 and 100~keV the $E_{\rm C}$ = 3$kT_{\rm e}$ relation does not 
work. This is due to the fact that the exponential cut-off changes the shape of 
the spectrum already in the energy band well below $kT_{\rm e}$, whereas the 
Comptonized spectrum cut-off is much sharper and appears close to the cut-off 
energy. In consequence, to recover the sharper cut-off the exponential cut-off 
models fitted to the spectra lead to the spectral models harder than those 
fitted with the Comptonization model. We tested this effect applying the 
{\tt PEXRAV} model to our spectra of NGC~4151. When we fix $E_{\rm C}$ at 
160~keV (equal to the $kT_{\rm e}$ = 53~keV multiplied by 3), the other 
parameters are strongly altered, with $\Gamma$ $\approx$ 1.55 and $R$ 
$\approx$ 0.45 instead of 1.75 and 0.33, respectively, found with the 
Comptonization model. Also the parameters of the partial absorber component 
were changed due to the need to adjust to the harder continuum.

Our sample, although not very large, is the largest Seyfert sample modelled 
with the complex physical model. We found that the plasma emitting in the 
X/$\gamma$ band is relatively cool, with only a small fraction of objects 
showing $kT_{\rm e}$ above 100~keV. This fact has to be taken into account in 
CXB modelling, where the template spectra are used to model the total emission 
from the Seyfert galaxies, dominating the AGN population. Contemporary 
synthesis models assume $E_{\rm C}$ in the 150--400~keV range. In panel (b) of 
Fig. \ref{xrbspec} we show several examples of the template spectra, compared 
with the Comptonization model corresponding to our median $kT_{\rm e}$, $y$ and 
$R$ values. The spectra used by \citet{Ballantyne2006} ($E_{\rm C}$ = 375~keV), 
\citet{Gilli2007} ($E_{\rm C}$ = 200~keV) and \citet{Ueda2014} ($E_{\rm C}$ = 
300~keV) are more concave below the peak due to the stronger Compton reflection 
assumed, $R$ = 1. All template spectra are less steep above the peak than our 
spectrum, with a considerable emission above 200~keV. The rightmost dash-dotted 
spectrum corresponding to the upper quartile of our $kT_{\rm e}$ distribution,
i.e. 105~keV, definitely overpredicts the high-energy emission and forms too 
broad peak to be consistent with the CXB spectrum. This is the result of the 
fact that our sample is not homogeneous. In future population studies the 
high-$kT_{\rm e}$ objects should be considered separately, provided that this
subsample will be enlarged.  

Any quantitative analysis of the CXB template spectra is out of the scope of the 
present paper. Definitely, a substantial change in the template model will 
affect the population synthesis model, altering its results in many respects.
If the low-temperature plasma is a common feature of not only local Seyferts,
there will be several consequences expected. The most important will be the need 
to reduce the expected population of the heavily absorbed, so-called 
Compton-thick (CT) AGN. These objects, with $\log(N_{H}) > 24$~cm$^{-2}$, can be 
detected only in the hard X-ray domain, where the sensitivity of the 
contemporary observatories remains too low for a detection of more distant 
sources. Current hard X-ray surveys place the fraction of CT AGN among all local 
AGN in the range between 0 and 20 per cent (e.g., 
\citealt{Beckmann2009,Burlon2011}). The cumulated emission from CT AGN forms a 
relatively narrow peak \citep{Gilli2007} at energy close to the CXB peak. 
Population synthesis models suggest the CT AGN fraction, corrected for the 
selection effect, in a wide range between 5 and 50 per cent \citep{Akylas2012}. 
If the typical Seyfert spectra are more peaked due to the lower high-energy 
cut-off than the template spectra used so far, there will be smaller population 
of CT AGN needed. The Compton reflection also contributes to the CXB peak. 
Although we found this component to be weaker than that assumed usually in the 
synthesis models, its role in forming the peak will be reinforced due to its 
narrower shape after reprocessing a primary continuum emission with lower 
temperature (cf. Fig. \ref{specs}).

The exponential cut-off template spectra shown in Fig. \ref{xrbspec} are less 
steep at high energy than the Comptonization model spectrum. This will affect 
additionally the synthesis model results, in particular the CXB peak, weakening 
it due to the emission from distant, high-redshift AGN. The fact that models 
with large $E_{\rm C}$ reproduce the CXB spectrum in a wide energy range 
indicates that there is some emission needed at high energy, in excess to that 
coming from the low-temperature Seyferts. Taking this into account more advanced 
studies of the AGN population should be possible, aiming at identifying the 
sources of the high-energy emission. These can be some rare, high-temperature, 
radio-quiet Seyfert galaxies like Akn~120 and NGC~2110. However, the main 
candidates for the sources dominating 100-300 keV emission are presumably 
radio-loud AGN. Observations show that the fraction of the radio-loud objects 
among all AGN is at most several per cent \citep{Best2009}. These objects are 
much stronger emitters above 100~keV than the Seyferts, being well fitted with 
the cut-off power-law models with $E_{\rm C}$ $>$ 300~keV, e.g., Cen A 
\citep{Beckmann2011} and 3C~273 \citep{Esposito2015}. Future population 
synthesis models can attempt to disentangle the CXB components related to the 
radio-quiet and radio-loud AGN.

\section{Conclusions}

We have analysed the summed, long-exposure time spectra of 28 Seyfert nuclei,
collected over the 2002--2012 period by the \integral/ISGRI detector. These 
hard X-ray/soft $\gamma$-ray data were fitted together with all contemporary 
medium X-ray spectra taken with the \xmm, \suzakue and \xtee satellites. We 
have applied a complex physical spectral model including a Comptonized 
continuum, a Compton reflection component and two absorption components 
corresponding to a neutral and to an ionized, partially covering absorption 
media. 
 
To take into account the spectral variability of the low-energy spectra, various 
options of the model were fitted, including variable spectral slope and variable 
absorption. As the result of a massive spectral fit campaign we have obtained a 
set of relatively precise average values of many physical parameters for the 
largest Seyfert sample so far studied with such an elaborate model.

A mean temperature of the electron plasma was found to be 26 $\leq kT_{\rm e} 
\leq$ 62~keV for a majority (19 among 28) of the studied objects. The rest of 
the sample exhibits mean temperatures 80 $\leq kT_{\rm e} \leq$ 360~keV. The 
distribution of the mean Compton parameter is bimodal, with a narrow peak
around $y$ = 1.1 and a broad peak around $y$ = 0.8, the latter corresponding to 
almost all high-temperature objects. Mean values of the photon index $\Gamma$, 
derived from the mean $y$ and $kT_{\rm e}$ values, occupy a narrow range between 
1.7 and 2.1, except for two objects. Compared to the previous analyses, we found 
much weaker Compton reflection strength, with a vast majority of the $R$ values 
well below 1, and only three objects showing $R$ $>$ 1.

Regardless the small size of subsamples we found very similar median values of 
the $kT_{\rm e}$, $y$ and $R$ parameters for the three types of Seyfert 
galaxies. On the other hand, radio-loud objects are characterized by higher 
$kT_{\rm e}$ and lower $y$ than the radio-quiet objects.

Our results are in agreement with an emerging paradigm of the hard X-ray source 
in Seyfert nuclei being compact and localized at the centre of the system, 
separated geometrically from the accretion disc. The narrow range of observed 
$y$ values points toward a quite similar plasma geometry within the 
low-temperature subsample. On the other hand, high-temperature objects show a 
larger range of the Compton parameter, related presumably to considerable 
differences of the system geometry and/or plasma heating and cooling mechanisms.

Reduced irradiation of the disc by the plasma photons and vice versa, together
with the low plasma temperatures observed, indicates that there should be 
another process, besides the disc emission, needed to explain an efficient 
plasma cooling. The most probable effect is the plasma synchrotron emission, 
within scenarios postulated by the recent hybrid Comptonization models.

We have found that the integrated X-ray and $\gamma$-ray luminosity provides 
a very good estimate of half of the total accretion luminosity of the Seyfert 
galaxies. This finding confirms a recent observation that the radiation emitted 
by the plasma region and the radiation of the accretion disc are approximately 
equal for these sources.

Compared with the results of the X-ray spectral analysis obtained for Galactic
BH binaries, our results support an analogy between BHBs in the hard state and
bright Seyfert galaxies. 

The low typical $kT_{\rm e}$ we found for the Seyferts' plasma implies that the 
template spectra adopted in AGN population synthesis models should be revised. 
The most important consequence of a shifted high-energy cut-off will be a
considerably smaller fraction of the Compton-thick AGN needed to explain the
peak of the CXB spectrum. For the same reason, a population of radio-loud AGN 
will be needed to balance the reduced emission of radio-quiet AGN in the 100
--300~keV band.

\section*{Acknowledgments}
PL and AAZ have been supported by the Polish NCN grants 2012/04/M/ST9/00780 and 
UMO-2014/13/B/ST9/00570. We thank the reviewer for his/her thorough review and 
highly appreciate the comments and suggestions, which significantly contributed 
to improving the quality of the publication. We used data from the High Energy 
Astrophysics Science Archive Research Center, and from the NASA/IPAC 
Extragalactic Database. We acknowledge the usage of the HyperLeda database 
(http://leda.univ-lyon1.fr). Most of the massive computations was done using the 
NCAC cluster in Toru{\'n}, excellently maintained by Jerzy Borkowski.

\bibliographystyle{mnras}
\bibliography{plnew}

\clearpage

%\afterpage{\clearpage}

%\begin{samepage}

\appendix

\section{Estimate of the black hole mass}
\label{appena}

There are various methods used to determine the mass of supermassive black holes 
residing in the center of normal and active galaxy nuclei. Each of those methods
has some limitations. For example, the reverberation mapping, probably the most 
robust technique besides that applying a water maser emission, can suffer from 
the radiation pressure affecting the BLR size estimate \citep{Marconi2008}.
Indeed, the reverberation-based mass estimates are usually smaller that the 
values derived with the other methods. In addition, reverberation mapping of the 
broad line region cannot be used for absorbed objects where this region is 
obscured. To reduce a role of the limitations of different methods we decided to
compute a weighted mean of the BH masses found in the literature together with
our estimates based on the stellar velocity dispersion data. 

\begin{table*}
\setlength{\tabcolsep}{1.0mm}
\centering
\caption{Various estimates of the masses of supermassive black holes for the 
sample of studied Seyfert galaxies. This work: $M_{\rm BH}$ derived from the 
average values of stellar velocity dispersion. ``Catalogue" is the second 
\integrale AGN catalog \citep{Beckmann2009} from which we adopted the codes of 
various methods applied for the $M_{\rm BH}$ estimate (see the text). The 
references quoted are: (1) \citealt{Winter2010}, (2) 
\citealt{Winter2009,Winter2012}, (3) \citealt{Vasudevan2010}, (4) 
\citealt{Bian2003}, (5) \citealt{Wang2007}, (6) \citealt{Khorunzhev2012}, (7) 
\citealt{Wang2009}, (8) \citealt{Peterson2004}, (9) \citealt{Markowitz2009}, 
(10) \citealt{Tazaki2010}, (11) \citealt{Denney2006}, (12) \citealt{Wandel2002}, 
(13) \citealt{Woo2010}, (14) \citealt{Stalin2011}, (15) 
\citealt{Marchesini2004}, (16) \citealt{Chatterjee2011}, (17) 
\citealt{Bentz2006}, (18) \citealt{Woo2002}, (19) \citealt{Uttley2005}, (20) 
\citealt{McHardy2005}, (21) \citealt{Middleton2008}, (22) \citealt{Bian2007}, 
(23) \citealt{Kuo2011}, (24) \citealt{Nicastro2003}, (25) \citealt{Zhou2005}, 
(26) \citealt{Graham2008}, (27) \citealt{Czerny2001}, (28) \citealt{Awaki2005}. 
Values in square brackets were not used (see the text).}
\label{masses}
\footnotesize{
\begin{tabular}{lcccccccccccc}
\hline
Object          & \multicolumn{12}{c}{$\log(M_{\rm BH})$ [$M_{\sun}$]} \\
\multicolumn{1}{r}{Reference} & This work & Catalogue & (1) & (2) & (3) & (4) & (5) & (5) & (6) & (6) & (7) & Others (reference) \\
\multicolumn{1}{r}{Method}    & S         &         & CL  & KM & KM  & CL  & LL  & SO  & KM  & LL  & LL  &                  \\
\hline
IC 4329A        & 8.41 & 7.00(R,8)  & ---  & 8.19 & 8.29   & 7.45 & 7.42 & 7.89 & 7.98 & ---  & ---  & 8.11(X+B,9) \\ % Peterson 2004
IGR J21247+5058 & ---  & ---        & [6.58] & ---  & ---    & ---  & ---  & ---  & ---  & ---  & ---  & 7.80(CL+LL,10) \\ 
GRS 1734-292    & [8.94] & 8.94(S)    & ---  & ---  & ---    & ---  & ---  & ---  & ---  & ---  & ---  & \\ % Iossif 
NGC 4593        & 8.11 & 6.99(R,11) & 7.83 & 8.29 & 7.51   & 7.12 & 7.22 & 6.55 & 7.98 & 7.10 & 7.12 & 6.91(B,12), 6.97(S,13), 6.882(R,29) \\ % Hicks & Malkan 2008 = Denney 2006
4U 0517+17      & 7.39 & ---        & ---  & ---  & ---    & ---  & 7.20 & 6.88 & ---  & 7.65 & 7.55 & 6.98(R,14) \\
Akn 120         & 8.26 & 8.18(R,8)  & ---  & 8.43 & 8.54   & 8.21 & 8.25 & 7.69 & ---  & ---  & 8.48 & 8.29(B,12), 8.15(S,13), 8.068(R,29) \\ % Peterson 2004
ESO 141-55      & 7.62 & 7.10(SO,5) & ---  & ---  & ---    & ---  & 7.91 & [7.10] & ---  & ---  & ---  & \\ % Middleton 2008 = Wang & Zhang 2007 
3C 111          & 7.10 & 9.56(B,15) & 8.54 & ---  & ---    & ---  & ---  &---   & ---  & 9.08 & 8.33 & 8.26(LL,16) \\ % Middleton 2008 = Grandi 2006 = Marchesini 2004 
\hline
NGC 4151        & 6.97 & 7.66(R,17) & 7.07 & 7.27 & ---    & 7.24 & 7.31 & 7.44 & 7.61 & 7.67 & 7.49 & 7.08(B,12), 7.64(S,13), 7.555(R,29) \\ % Hicks & Malkan = Bentz 2006
MCG+8-11-11     & 6.37 & 8.06(SO,5) & 8.07 & ---  & 8.17   & 7.18 & 7.89 & [8.06] & ---  & ---  & 7.60 & \\ % Middleton 2008 = Wang & Zhang 2007
Mrk 509         & ---  & 8.16(R,8)  & ---  & 8.26 & 8.56   & 7.70 & 7.66 & 7.80 & 8.15 & 8.22 & 7.47 & 7.98(B,13), 9.049(R,29) \\ % Peterson 2004
4U 1344-60      & 7.89 & ---        & ---  & ---  & 7.44   & ---  & ---  & ---  & ---  & ---  & ---  & \\
3C 120          & 7.86 & 7.74(R,8)  & ---  & 8.23 & 8.35   & 6.85 & ---  & ---  & ---  & ---  & ---  & 8.13(S,18), 7.49(B,12), 7.72(S,13), 7.745(R,29) \\ % Peterson 2004
NGC 6814        & 7.12 & 7.08(CL,4) & ---  & 7.78 & ---    & 6.95 & ---  & ---  & 7.53 & 6.91 & 6.95 & 7.28(CL,18), 7.25(S,13), 7.038(R,29) \\ % Hicks & Malkan 2008 = Wandel 2002
3C 390.3        & 8.64 & 8.46(R,8)  & ---  & 8.19 & 8.35   & 8.51 & ---  & ---  & ---  & 9.61 & 9.13 & 8.59(B,12), 8.44(S,13), 8.638(R,29) \\ % Peterson 2004
MCG-6-30-15     & 6.87 & [6.65](X,19) & ---  & 6.91 & 7.25   & 6.19 & ---  & ---  & 7.26 & ---  & ---  & 6.46(X,20), 6.81(SO,21)  \\ % Uttley & McHardy 2005
\hline
NGC 4388        & 6.90 & 7.22(S,22) & ---  & 8.20 & 7.07   & ---  & ---  & 8.16 & 7.20 & ---  & ---  & 6.93(M,23), 8.54(B,24) \\ % Bian & Gu 2007
NGC 2110        & 8.55 & 8.30(S,4)  & ---  & 7.92 & 7.40   & ---  & ---  & 7.96 & ---  & ---  & ---  & \\ % Woo & Urry 2002
NGC 4507        & 7.63 & 7.58(S,22) & ---  & 8.04 & 7.70   & ---  & ---  & 6.42 & 7.81 & ---  & ---  & 8.26(B,24) \\ % Middleton 2008 = Bian & Gu 2007
MCG-5-23-16     & 7.94 & 6.29(SO,5) & ---  & 7.24 & ---    & ---  & ---  & [6.29] & 7.28 & ---  & ---  & 7.85(N,25) \\ % Middleton 2008 = Wang & Zhang 2007
NGC 5506        & 6.88 & 6.65(S,22) & ---  & 7.36 & 7.67   & ---  & ---  & 7.46 & 7.68 & ---  & ---  & 8.00(B,24) \\ % Bian & Gu 2007
Cygnus A        & ---  & 9.41(K,26) & ---  & ---  & ---    & ---  & ---  & ---  & ---  & ---  & ---  & \\ % 3C 405.0 (9.32 Marconi 2004 - no data), Graham 2008
NGC 5252        & 8.09 & 8.98(K,26) & ---  & 8.32 & ---    & ---  & ---  & 7.56 & ---  & ---  & ---  & 8.56(B,24) \\ % Graham 2008
ESO 103-35      & 7.15 & 7.14(X,27) & ---  & 7.32 & 7.30   & ---  & ---  & ---  & ---  & ---  & ---  & \\ % Middleton 2008 = Czerny 2001
NGC 788         & 7.53 & 7.51(S,22) & ---  & 8.18 & 7.39   & ---  & ---  & 6.04 & 7.92 & ---  & ---  & \\ % Bian & Gu 2007
NGC 6300        & 6.94 & 5.45(X,28) & ---  & ---  & $>$6.7 & ---  & ---  & ---  & 7.59 & ---  & ---  & \\ % Middleton 2008 = Awaki 2005 
NGC 1142        & 8.26 & [9.36](KM,2) & ---  & 9.11 & 7.98   & ---  & ---  & ---  & 8.48 & ---  & ---  & \\ % Winter 2009
LEDA 170194     & ---  & [8.88](KM,2) & ---  & 8.58 & ---    & ---  & ---  & ---  & ---  & ---  & ---  & \\ % IGR J12391-1612, Winter 2009
\hline
\end{tabular}                                                                   
}                                                                               
\end{table*}

The numbers used for the mean computation are shown in the Table \ref{masses}.
In this Table we adopted method's symbols defined in the second \integrale AGN
catalogue \citep{Beckmann2009}: `R' - reverberation mapping, `SO' - indirect
estimate of $\sigma_{\star}$ from the O~III line width, `CL' - size of the 
BLR from the continnum luminosity, `LL' - size of the broad line region from 
the H~$\beta$ line luminosity, `KM' - K-band stellar magnitude, `S' - a direct 
measure of the stellar velocity dispersion, `X' - X-ray power spectral density, 
`M' - water maser, `B' - bulge luminosity, `N' - narrow line region size 
from the H~$\beta$ line.

Our estimate uses the average values of stellar velocity dispersion 
$\sigma_{\star}$ taken from the Hyper Leda database \citep{Paturel2003}, except 
for MCG--6-30-15, NGC~4507, NGC~6300 and NGC~1142 where we used $\sigma_{\star}$ 
values from \citealt{Garcia2005} and for NGC~5506 and ESO~103-55 from 
\citealt{Gu2006}. The velocity dispersions were converted into $M_{\rm BH}$ 
values using the \citealt{Tremaine2002} formula. We tried also a more recent 
formula of \citealt{Gultekin2009} but the logarithms of BH masses were almost 
the same (difference $\leq 0.1$) as for the \citealt{Tremaine2002} equation. 
There are several other estimates using method `S' in the Table \ref{masses}, 
however, we have checked that there were other $\sigma_{\star}$ data used. In 
the second column of Table \ref{masses} there are literature values compiled in 
the \citealt{Beckmann2009} catalogue, with the reference given. The other 
literature numbers are taken from the papers listed in the caption of Table 
\ref{masses}. We have checked those data to avoid a double use of the same 
value. If for a given method there are at least two numbers used for the mean 
computation, they are based on either a different approach for the same 
observational data or a different data set. For example, \citealt{Vasudevan2010} 
and \citealt{Winter2012} estimates, based on the K-band stellar magnitudes from 
the Two Micron All Sky Survey (2MASS), were obtained with different corrections 
and different formulas to convert them into BH mass values. Thus, we treated 
these two estimates as independent measures.

Black hole masses determined with the reverberation mapping shown in Table 
\ref{masses} were taken from \citealt{Peterson2004} unless there was a newer,
more precise result published. In the case of \citealt{Winter2012} we quote the 
original numbers from the paper for Sy~1 and Sy~1.5 objects, whereas for Sy~2 
we used \citealt{Winter2009} data and a correction found in 
\citealt{Winter2010}. Several values shown in the Table \ref{masses}, put into
square brackets, were skipped during the mean computation. These are three 
numbers from \citealt{Wang2007} included in the \citealt{Beckmann2009} data,
two catalogue values for NGC~1142 and LEDA~170194 where the \citealt{Winter2009}
values were corrected by us (fifth column of the Table \ref{masses}) and our 
estimate for GRS~1734-292 included already in the \citealt{Beckmann2009} 
catalogue. We also excluded the \citealt{Winter2010} value for IGR~J21247+5058 
because their extinction correction for the continuum emission at 5100~\AA{} 
seems too large \citep{Tazaki2010}. At last, the `Catalogue' value for 
MCG--6-30-15 was excluded too, because this is a mean value for different 
methods computed by \citealt{Uttley2005}. In this case we used the original 
\citealt{Uttley2005} values derived with the X-ray variability method, quoted 
in the last column of Table \ref{masses}.

The weighted mean of the BH masses was computed for the logarithm values, with 
the uncertainty for different methods expressed in dex units. For the 
`Catalogue' we used the original errors quoted in \citealt{Beckmann2009}. For 
the other data we applied the uncertainty estimates from the middle of the 
uncertainty limits suggested in \citealt{Beckmann2009}. Thus, the errors of 
$\log(M_{\rm BH}$) adopted by us were: 0.1~dex for method `M', 0.2~dex for 
methods `K' and `R', 0.3~dex for method `S', 0.5~dex for methods `B', `CL', 
`KM', `LL', and 0.7~dex for methods `N' and `X'. As a check we tried also an 
arithmetic mean which produced the results quite similar to the weighted mean.

\newpage

\section{Observation log}
\label{appenb}

\begin{table}
\centering                                                                      
\caption{Details of the observations of the Seyferts sample done with \integrale 
and other X-ray satellites. The last column shows the spectrum code within a 
spectral set used for a given object: I - \integral/ISGRI, X - \xmm/EPIC pn, S - 
\suzaku/XIS, R - \xte/PCA.}   
\footnotesize                                                         
\label{xspectra}
\begin{tabular}{@{}lrrrc@{}}                                                  
\hline
OBS ID & Start date & End date & Exposure [ks] & Spectrum \\    
\hline
\multicolumn{5}{c}{IC 4329A} \\
0030--0776  & 2002-12-21 & 2009-05-22 &  1774 & I \\
0101040401  & 2001-01-31 & 2001-01-31 &  13.9 & X2 \\
0147440101  & 2003-08-06 & 2003-08-07 & 136.0 & X1 \\
702113010   & 2007-08-01 & 2007-08-01 &  25.4 & S1 \\
702113020   & 2007-08-06 & 2007-08-06 &  30.6 & S2 \\
702113030   & 2007-08-11 & 2007-08-11 &  26.9 & S3 \\
702113040   & 2007-08-16 & 2007-08-16 &  24.2 & S4 \\
702113050   & 2007-08-20 & 2007-08-17 &  24.0 & S5 \\
40153-01    & 2001-08-21 & 2001-08-27 &  52.6 & R1 \\ 
80152-03/04 & 2003-04-08 & 2004-02-26 & 140.0 & R3 \\ 
70148-01    & 2003-08-20 & 2003-08-23 &  60.8 & R2 \\  
90154-01    & 2004-03-01 & 2007-07-17 &  58.4 & R4 \\ 
91138-01    & 2005-03-07 & 2007-08-07 &  57.6 & R5 \\ 
92108-01    & 2006-03-05 & 2007-06-26 &  65.1 & R6 \\ 
\hline
\multicolumn{5}{c}{IGR J21247+5058} \\
0023--0806  & 2002-12-21 & 2009-05-22 &  5656 & I \\
0306320101  & 2005-05-05 & 2005-05-05 &  11.8 & X2 \\
0306320201  & 2005-11-06 & 2005-11-07 &  27.8 & X1 \\
702027010   & 2007-04-16 & 2007-04-17 &  85.0 & S1 \\
\hline
\multicolumn{5}{c}{GRS 1734-292} \\
0046--0849  & 2003-02-28 & 2009-09-28 & 15214 & I \\
0550451501  & 2009-02-26 & 2009-02-26 &  17.9 & X1 \\
\hline
\multicolumn{5}{c}{NGC 4593} \\
0028--0880  & 2003-01-05 & 2009-12-29 &  2962 & I \\
0109970101  & 2000-07-02 & 2000-07-02 &  28.1 & X2 \\
0059830101  & 2002-06-23 & 2002-06-24 &  53.2 & X1 \\
702040010   & 2007-12-15 & 2007-12-16 & 118.8 & S1 \\
70145-02/03 & 2002-06-25 & 2002-07-10 & 174.3 & R6 \\ 
90160-02    & 2004-02-28 & 2006-02-09 & 151.7 & R3 \\ 
91140-04    & 2005-11-28 & 2006-01-31 & 174.1 & R1 \\ 
92113-02    & 2006-02-13 & 2007-06-28 & 154.7 & R2 \\ 
93127-05    & 2007-07-13 & 2008-12-25 & 126.5 & R4 \\ 
94127-05    & 2008-12-27 & 2009-12-30 & 121.0 & R5 \\ 
\hline
\multicolumn{5}{c}{4U 0517+17} \\
0039--1328  & 2003-02-16 & 2013-08-29 &  2399 & I \\
0502090501  & 2007-08-21 & 2007-08-22 &  61.9 & X1 \\
707029010   & 2013-02-28 & 2013-02-29 &  46.0 & S1 \\
\hline
\multicolumn{5}{c}{Akn 120} \\
0041--0839  & 2003-02-14 & 2009-08-28 &  2209 & I \\
0147190101  & 2003-08-24 & 2003-08-25 & 112.1 & X1 \\
702014010   & 2007-04-01 & 2007-04-02 & 100.9 & S1 \\
80160-03    & 2003-08-24 & 2006-12-28 & 113.7 & R1 \\
\hline
\multicolumn{5}{c}{ESO 141-55} \\
0050--1262  & 2003-03-13 & 2013-02-13 &   979 & I \\
0503750301  & 2007-10-09 & 2007-10-10 &  30.5 & X3 \\
0503750401  & 2007-10-10 & 2007-10-10 &  27.3 & X4 \\
0503750501  & 2007-10-12 & 2007-10-13 &  75.9 & X2 \\
0503750101  & 2007-10-30 & 2007-10-31 &  79.5 & X1 \\
\hline
\multicolumn{5}{c}{3C 111} \\
0047--0837  & 2003-03-03 & 2009-08-23 &  2676 & I \\
0065940101  & 2001-03-14 & 2001-03-15 &  44.8 & X2 \\
0552180101  & 2009-02-15 & 2009-02-17 & 124.6 & X1 \\ 
703034010   & 2008-08-22 & 2008-08-23 & 122.4 & S1 \\
90152-01    & 2004-03-01 & 2006-03-09 & 136.2 & R1 \\ 
91146-01    & 2005-03-31 & 2006-09-20 & 123.9 & R2 \\ 
\hline
\end{tabular} 
\end{table}

\begin{table}
\centering                                                                      
\contcaption{}   
\footnotesize                                                         
\begin{tabular}{@{}lrrrc@{}}                                                  
\hline
OBS ID & Start date & End date & Exposure [ks] & Spectrum \\    
\hline
92102-01/04 & 2006-08-13 & 2007-05-04 & 221.0 & R3 \\ 
93137-01/03 & 2007-06-29 & 2008-12-25 & 136.3 & R4 \\ 
94137-01    & 2008-12-28 & 2009-12-30 &  58.6 & R5 \\ 
\hline
\multicolumn{5}{c}{NGC 4151} \\
0036--1179  & 2003-01-29 & 2012-06-11 &  3236 & I \\
0112310501  & 2000-12-21 & 2000-12-21 &   7.9 & X1 \\
0112310101  & 2000-12-21 & 2000-12-22 &  33.0 & X1 \\
0112830501  & 2000-12-22 & 2000-12-22 &  23.1 & X1 \\
0112830201  & 2000-12-22 & 2000-12-23 &  62.2 & X1 \\
0112830601  & 2000-12-23 & 2000-12-23 &   5.2 & X1 \\
0143500101  & 2003-05-25 & 2003-05-25 &  19.0 & X2 \\
0143500201  & 2003-05-26 & 2003-05-27 &  18.9 & X2 \\
0143500301  & 2003-05-27 & 2003-05-27 &  19.0 & X3 \\
0402660101  & 2006-05-16 & 2006-05-16 &  40.4 & X4 \\
0402660201  & 2006-11-29 & 2006-11-30 &  52.8 & X5 \\
0657840101  & 2011-05-11 & 2011-05-11 &  16.2 & X6 \\
0657840201  & 2011-06-12 & 2011-06-12 &  16.3 & X7 \\
0657840301  & 2011-11-25 & 2011-11-25 &  10.9 & X8 \\
0657840401  & 2011-12-09 & 2011-12-09 &  10.2 & X9 \\
0679780101  & 2012-05-13 & 2012-05-13 &   9.9 & X10 \\
0679780201  & 2012-06-10 & 2012-06-10 &   8.7 & X11 \\
701034010   & 2006-12-18 & 2006-12-20 & 125.0 & S1 \\
906006010   & 2011-11-17 & 2011-11-18 &  61.7 & S2 \\
906006020   & 2011-12-18 & 2011-12-19 &  60.6 & S3 \\       
\hline
\multicolumn{5}{c}{MCG+8-11-11} \\
0218--0966  & 2004-08-16 & 2010-09-12 &  1404 & I \\
0201930201  & 2004-04-09 & 2004-04-10 &  38.4 & X1 \\
702112010   & 2007-09-17 & 2007-09-18 &  98.7 & S1 \\
00018       & 1996-01-22 & 1996-01-22 &  16.6 & R1 \\ 
\hline
\multicolumn{5}{c}{Mrk 509} \\
0066--0867  & 2003-04-29 & 2009-11-20 &  1371 & I \\ 
0306090201  & 2005-10-18 & 2005-10-19 &  85.9 & X3 \\
0306090301  & 2005-10-20 & 2005-10-20 &  47.1 & X5 \\
0306090401  & 2006-04-25 & 2006-04-25 &  69.9 & X4 \\
0601390201  & 2009-10-15 & 2009-10-16 &  60.9 & X1 \\
0601390301  & 2009-10-19 & 2009-10-20 &  63.8 & X1 \\
0601390501  & 2009-10-29 & 2009-10-30 &  60.9 & X6 \\
0601390601  & 2009-11-02 & 2009-11-03 &  62.8 & X1 \\
0601390701  & 2009-11-06 & 2009-11-07 &  63.1 & X2 \\
0601390901  & 2009-11-14 & 2009-11-15 &  60.9 & X2 \\
0601391001  & 2009-11-18 & 2009-11-18 &  65.5 & X1 \\
0601391101  & 2009-11-20 & 2009-11-21 &  62.8 & X2 \\
701093010   & 2006-04-25 & 2006-04-26 &  24.6 & S1 \\
701093020   & 2006-10-14 & 2006-10-14 &  25.9 & S2 \\
701093030   & 2006-11-15 & 2006-11-15 &  24.4 & S3 \\
701093040   & 2006-11-27 & 2006-11-27 &  33.1 & S4 \\
80157-01    & 2003-03-28 & 2006-07-29 & 190.6 & R1 \\ 
90147-01    & 2004-02-29 & 2006-07-25 & 169.6 & R2 \\ 
91129-01    & 2005-03-04 & 2006-07-19 &  79.2 & R3 \\ 
\hline
\multicolumn{5}{c}{4U 1344-60} \\
0030--0850  & 2003-01-11 & 2009-09-30 &  5781 & I \\
0092140101  & 2001-08-25 & 2001-08-25 &  36.9 & X1 \\
705058010   & 2011-01-11 & 2011-01-13 &  93.9 & S1 \\
\hline
\multicolumn{5}{c}{3C 120} \\
0040--0839  & 2003-02-25 & 2009-08-28 &  1721 & I \\
0109131101  & 2002-09-06 & 2002-09-06 &  12.6 & X2 \\
0152840101  & 2003-08-26 & 2003-08-27 & 133.8 & X1 \\
700001010   & 2006-02-09 & 2006-02-09 &  41.9 & S1 \\
700001020   & 2006-02-16 & 2006-02-17 &  41.6 & S2 \\
700001030   & 2006-02-23 & 2006-02-24 &  40.9 & S3 \\
700001040   & 2006-03-02 & 2006-03-03 &  40.9 & S4 \\
\hline
\end{tabular} 
\end{table}

\begin{table}
\centering                                                                      
\contcaption{}   
\footnotesize                                                         
%\label{xspectra}
\begin{tabular}{@{}lrrrc@{}}                                                  
\hline
OBS ID & Start date & End date & Exposure [ks] & Spectrum \\    
\hline
70164-01    & 2002-03-01 & 2003-04-27 & 139.3 & R1 \\ 
80175-01    & 2003-04-29 & 2006-03-02 & 287.1 & R2 \\ 
90152-02    & 2004-06-24 & 2007-05-01 & 234.9 & R3 \\ 
91146-02    & 2005-06-24 & 2007-02-27 & 204.1 & R4 \\ 
92102-03    & 2006-11-15 & 2007-01-13 & 183.5 & R5 \\ 
\hline
\multicolumn{5}{c}{NGC 6814} \\
0049--1041  & 2003-03-10 & 2011-04-25 &  3664 & I \\ 
0550451801  & 2009-04-22 & 2009-04-22 &  32.3 & X1 \\
706032010   & 2011-11-02 & 2011-11-03 &  42.1 & S1 \\
30015-02    & 1998-05-18 & 1998-08-21 &  18.6 & R1 \\ 
\hline
\multicolumn{5}{c}{3C 390.3} \\
0153--0829  & 2004-01-15 & 2009-07-30 &  1794 & I \\
0203720201  & 2004-10-08 & 2004-10-09 &  70.4 & X1 \\
0203720301  & 2004-10-17 & 2004-10-17 &  52.8 & X2 \\
708034010   & 2013-05-24 & 2013-05-25 & 100.4 & S1 \\
50178-01    & 2000-03-03 & 2001-02-23 & 165.2 & R1 \\ 
90130-01    & 2005-01-12 & 2005-01-14 &  81.4 & R2 \\ 
\hline
\multicolumn{5}{c}{MCG-6-30-15} \\
0030--0776  & 2003-01-11 & 2009-02-21 &  2174 & I \\
0111570101  & 2000-07-11 & 2000-07-11 &  46.4 & X1 \\
0111570201  & 2000-07-11 & 2000-07-12 &  66.2 & X2 \\
0029740101  & 2001-07-31 & 2001-08-01 &  89.4 & X3 \\
0029740701  & 2001-08-02 & 2001-08-02 & 129.4 & X4 \\ 
0029740801  & 2001-08-04 & 2001-08-05 & 130.5 & X5 \\            
100004010   & 2005-08-17 & 2005-08-17 &  46.7 & S1 \\             
700007010   & 2006-01-09 & 2006-01-11 & 143.2 & S2 \\
700007020   & 2006-01-23 & 2006-01-24 &  98.5 & S3 \\
700007030   & 2006-01-27 & 2006-01-29 &  96.7 & S4 \\
80154-03    & 2003-06-08 & 2005-02-18 & 125.0 & R1 \\ 
91140-10    & 2005-03-07 & 2007-05-05 &  59.0 & R2 \\  
92113-06    & 2006-03-05 & 2007-06-26 &  68.8 & R3 \\ 
93127-15/17 & 2007-05-19 & 2009-02-27 & 166.3 & R4 \\ 
94341-02    & 2009-03-03 & 2009-12-28 &  48.8 & R5 \\ 
\hline
\multicolumn{5}{c}{NGC 4388} \\
0028--1012  & 2003-01-05 & 2011-01-27 &  4815 & I \\
0110930701  & 2002-12-12 & 2002-12-12 &  12.0 & X2 \\
0675140101  & 2011-06-17 & 2011-06-18 &  61.4 & X1 \\
800017010   & 2005-12-24 & 2005-12-25 & 123.6 & S1 \\
\hline
\multicolumn{5}{c}{NGC 2110} \\
0051--1035  & 2003-03-16 & 2011-04-06 &  1736 & I \\
0145670101  & 2003-03-05 & 2003-03-06 &  59.6 & X1 \\
100024010   & 2005-09-16 & 2005-09-17 & 101.7 & S1 \\
30233-01    & 1997-12-07 & 1998-08-31 & 166.8 & R1 \\ 
80159-01    & 2003-03-05 & 2003-03-06 &  13.7 & R2 \\ 
\hline
\multicolumn{5}{c}{NGC 4507} \\
0048--0776  & 2003-03-07 & 2009-02-21 &  1600 & I \\
0006220201  & 2001-01-04 & 2001-01-05 &  46.2 & X1 \\
0653870201  & 2010-06-24 & 2010-06-24 &  19.9 & X2 \\	
0653870301  & 2010-07-03 & 2010-07-03 &  16.9 & X3 \\	
0653870401  & 2010-07-13 & 2010-07-14 &  16.9 & X4 \\
0653870501  & 2010-07-23 & 2010-07-23 &  16.9 & X5 \\
702048010   & 2007-12-20 & 2007-12-21 & 103.6 & S1 \\
10325-02    & 1996-02-24 & 1996-03-10 & 128.6 & R1 \\ 
80162-02-01 & 2003-06-25 & 2003-06-26 &   8.2 & R2 \\ 
\hline
\end{tabular} 
\end{table}

\begin{table}
\centering                                                                      
\contcaption{}   
\footnotesize                                                         
%\label{xspectra}
\begin{tabular}{@{}lrrrc@{}}                                                  
\hline
OBS ID & Start date & End date & Exposure [ks] & Spectrum \\    
\hline
\multicolumn{5}{c}{MCG-05-23-016} \\
0028--0770  & 2003-01-05 & 2009-02-03 &   839 & I \\
0112830401  & 2001-12-01 & 2001-12-02 &  24.9 & X2 \\
0302850201  & 2005-12-08 & 2005-12-10 & 131.7 & X1 \\
700002010   & 2005-12-07 & 2005-12-09 &  95.7 & S1 \\
10307-01    & 1996-04-24 & 1997-01-10 & 145.8 & R1 \\ 
91703-14    & 2005-12-09 & 2005-12-09 &  12.8 & R2 \\ 
\hline
\multicolumn{5}{c}{NGC 5506} \\
0095--0768  & 2003-07-24 & 2009-01-27 &  1479 & I \\
0201830201  & 2004-07-11 & 2004-07-11 &  21.6 & X1 \\
0201830301  & 2004-07-14 & 2004-07-15 &  20.4 & X2 \\
0201830401  & 2004-07-22 & 2004-07-22 &  22.0 & X3 \\
0201830501  & 2004-08-07 & 2004-08-07 &  20.4 & X4 \\
0554170201  & 2008-07-27 & 2008-07-28 &  90.9 & X5 \\
0554170101  & 2009-01-02 & 2009-01-04 &  88.9 & X6 \\
701030010   & 2006-08-08 & 2006-08-09 &  47.8 & S1 \\
701030020   & 2006-08-11 & 2006-08-12 &  53.3 & S2 \\
701030030   & 2007-01-31 & 2007-01-31 &  57.4 & S3 \\
60133-02    & 2001-03-02 & 2002-05-20 & 136.9 & R1 \\ 
90145-01    & 2004-07-11 & 2004-08-08 &  30.5 & R2 \\ 
\hline
\multicolumn{5}{c}{Cygnus A} \\
0022--0806  & 2002-12-18 & 2009-05-22 &  5614 & I \\
0302800101  & 2005-10-14 & 2005-10-14 &  22.5 & X1 \\
0302800201  & 2005-10-16 & 2005-10-16 &  18.8 & X2 \\
803050010   & 2008-11-15 & 2008-11-16 &  44.7 & S1 \\
\hline
\multicolumn{5}{c}{NGC 5252} \\
0028--1256  & 2003-01-05 & 2013-01-26 &  3674 & I \\
0152940101  & 2003-07-18 & 2003-07-18 &  67.3 & X1 \\
707028010   & 2012-12-26 & 2012-12-27 &  50.3 & S1 \\
\hline
\multicolumn{5}{c}{ESO 103-35} \\
0047--1030  & 2003-03-03 & 2011-03-21 &  1090 & I \\
0109130601  & 2002-03-15 & 2002-03-15 &  12.9 & X1 \\
703031010   & 2008-10-22 & 2008-10-23 &  91.4 & S1 \\
20324-01    & 1997-04-11 & 1997-11-14 & 131.6 & R1 \\ 
\hline
\multicolumn{5}{c}{NGC 788} \\
0098--0701  & 2003-08-02 & 2008-07-11 &  2136 & I \\
0601740201  & 2010-01-15 & 2010-01-16 &  35.4 & X1 \\
703032010   & 2008-07-13 & 2008-07-13 &  45.9 & S1 \\
\hline
\multicolumn{5}{c}{NGC 6300} \\
0037--0850  & 2003-02-01 & 2009-10-01 &  3580 & I \\
0059770101  & 2001-03-02 & 2001-03-02 &  53.2 & X1 \\
702049010   & 2007-10-17 & 2007-10-18 &  82.6 & S1 \\
10321-01/02 & 1997-02-14 & 1997-02-20 &  26.0 & R1 \\ 
\hline
\multicolumn{5}{c}{NGC 1142} \\
0098--0701  & 2003-08-02 & 2008-07-11 &  1629 & I \\
0312190401  & 2006-01-28 & 2006-01-28 &  11.9 & X1 \\
701013010   & 2007-01-23 & 2007-01-24 & 101.6 & S1 \\
702079010   & 2007-07-21 & 2007-07-21 &  40.6 & S2 \\
\hline
\multicolumn{5}{c}{LEDA 170194} \\
0028--0768  & 2003-01-06 & 2009-01-27 &  1135 & I \\
0670390301  & 2012-01-21 & 2012-01-21 &  16.9 & X1 \\
703007010   & 2008-12-18 & 2008-12-20 & 130.3 & S1 \\
\hline
\end{tabular} 
\end{table}

\clearpage

\section{ISGRI spectra}
\label{appenc}

\begin{figure*}
\includegraphics[width=\textwidth]{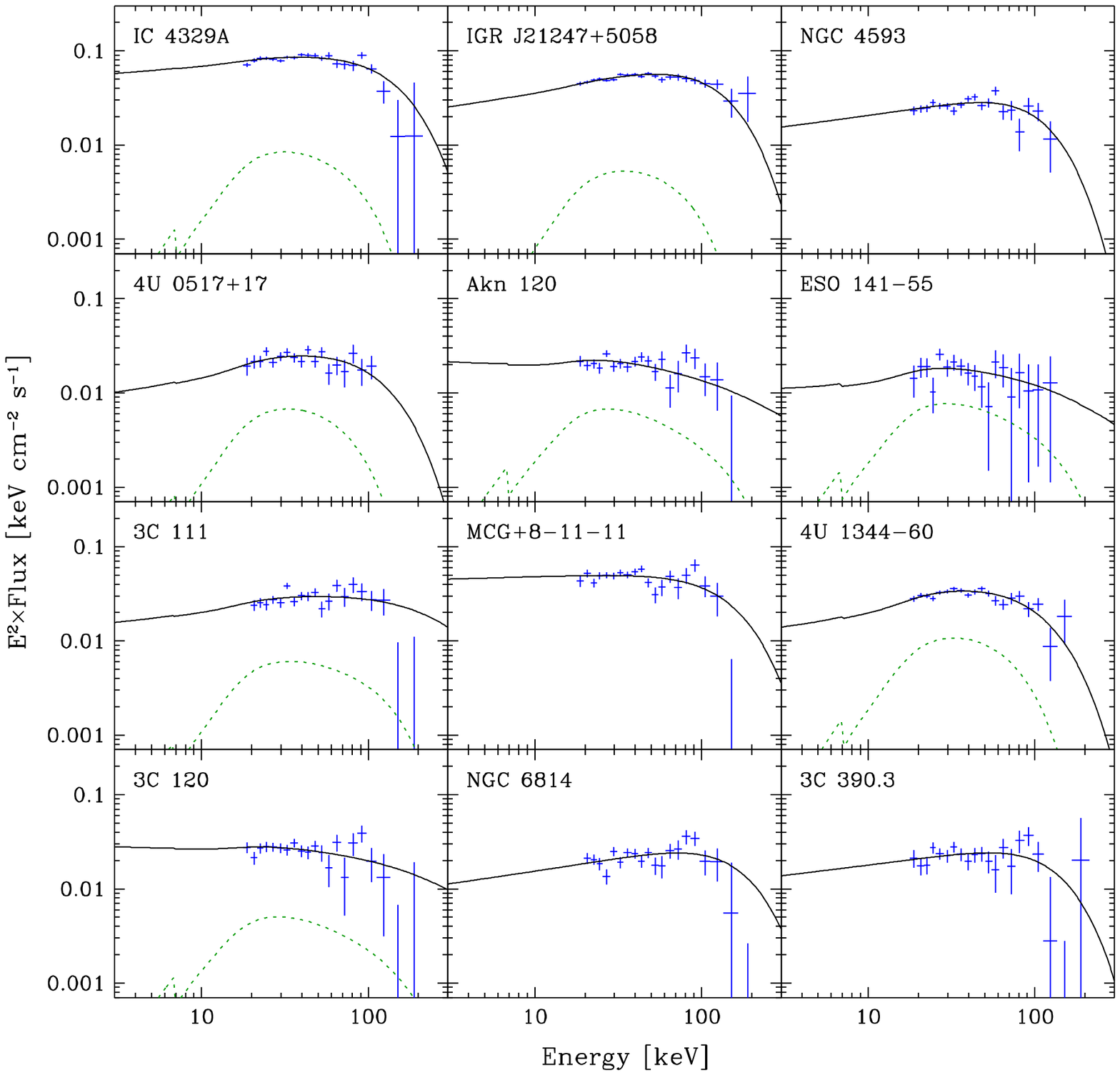}
\caption{Unfolded ISGRI (blue) spectra of Type 1 and Type 1.5 Seyfert galaxies.
Solid line - an unabsorbed Comptonization plus reflection model, dotted line - 
the reflection component, if present.}
\label{syspec1}
\end{figure*}

\begin{figure*}
\includegraphics[width=\textwidth]{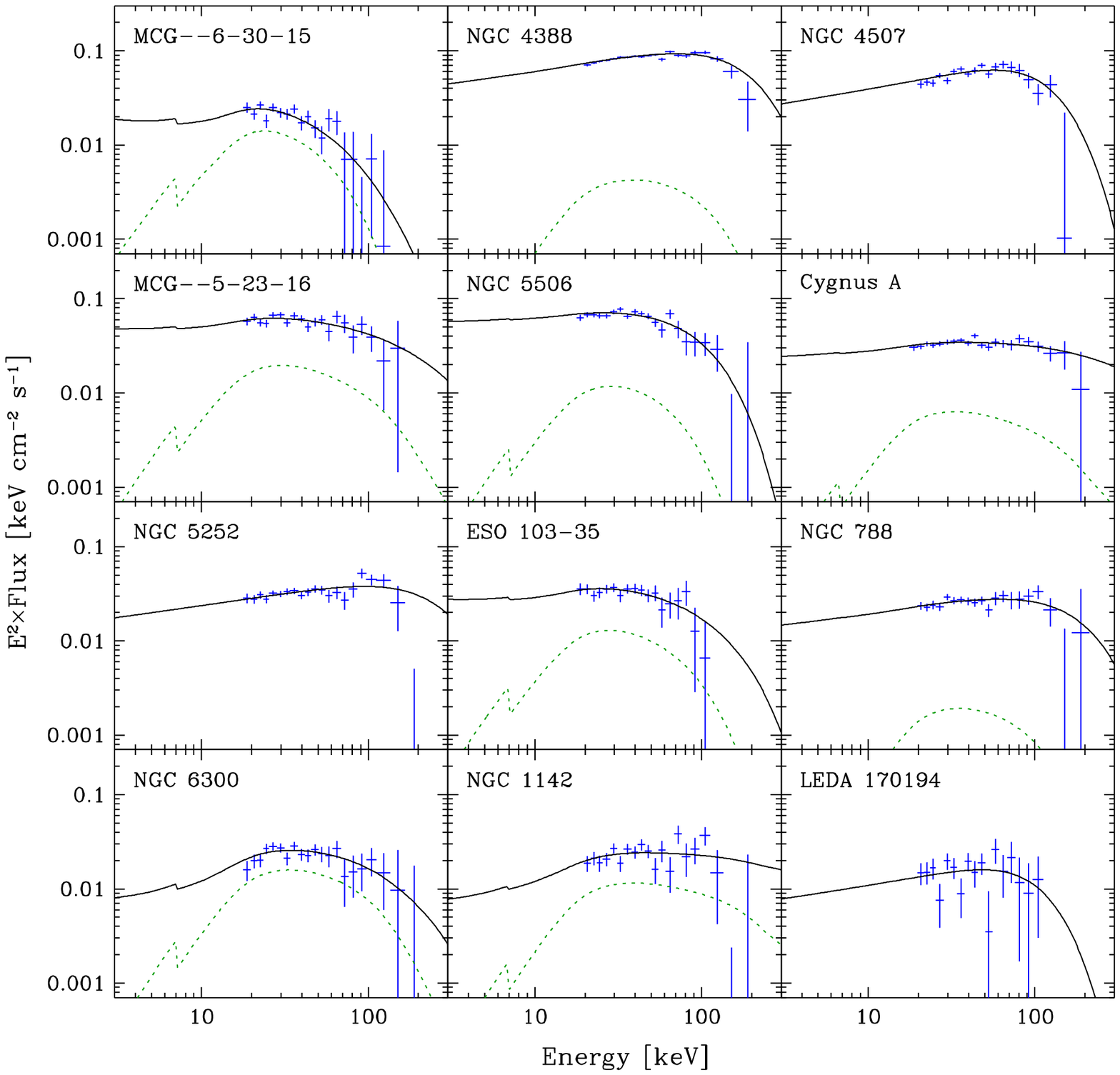}
\caption{Unfolded ISGRI (blue) spectra of MCG--6-30-15 and Type 2 Seyfert
galaxies. Solid line - an unabsorbed Comptonization plus reflection model, 
dotted line - the reflection component, if present.}
\label{syspec2}
\end{figure*}

In Figures \ref{syspec1} and \ref{syspec2} we present the ISGRI spectra of all 
objects, except for the spectra of NGC~4151, GRS~1734-292, Mrk~509 and NGC~2110,
presented in Sec. \ref{modopt}. The plotted models correspond to the 
best-fitting models, listed in Table \ref{mainres}. 

% Don't change these lines
\bsp    % typesetting comment

\label{lastpage}

\end{document}